\documentclass[fleqn,usenatbib]{mnras}

\usepackage[T1]{fontenc}
\usepackage{ae,aecompl}

\usepackage{graphicx}	% Including figure files
\usepackage{amsmath}	% Advanced maths commands
\usepackage{amssymb}	% Extra maths symbols
\usepackage{newtxtext,newtxmath}

\title[A search for young exoplanets in \textit{TESS} Sectors 1-5]{YOUNG Star detrending for Transiting Exoplanet Recovery (YOUNGSTER) - I. A search for young exoplanets in Sectors 1-5 of the \textit{TESS} Full-Frame-Images}
\author[M. P. Battley et al.]{
Matthew P. Battley,$^{1,2}$\thanks{E-mail: Matthew.Battley@warwick.ac.uk}
Don Pollacco,$^{1,2}$
David J. Armstrong$^{1,2}$
\\
$^{1}$Dept. of Physics, University of Warwick, Gibbet Hill Road, Coventry CV4 7AL, UK\\
$^{2}$Centre for Exoplanets and Habitability, University of Warwick, Gibbet Hill Road, Coventry CV4 7AL, UK\\
}

\date{Accepted XXX. Received YYY; in original form ZZZ}

\pubyear{2020}

\begin{document}
\label{firstpage}
\pagerange{\pageref{firstpage}--\pageref{lastpage}}
\maketitle

\begin{abstract}
%This is a simple template for authors to write new MNRAS papers.
%The abstract should briefly describe the aims, methods, and main results of the paper.
%It should be a single paragraph not more than 250 words (200 words for Letters).
%No references should appear in the abstract.
Young (<1Gyr) exoplanets represent a critically important area of exoplanet research, as they offer the opportunity to learn about the formation and early dynamic history of exoplanetary systems. However, finding young exoplanets is significantly complicated by the fast rotation and complex activity of their young host stars, which are often not well handled by state-of-the-art automatic pipelines. This work presents an alternative LOWESS-based pipeline focused specifically on detrending young stellar light-curves from the 30min-cadence Full Frame Images (FFIs) produced by the \textit{Transiting Exoplanet Survey Satellite (TESS)}, and includes improvements such as automatic peak-cutting of stellar variability and interpolation over masked transits to improve periodogram visibility and returned transit shapes. This work presents the details of the developed pipeline, along with initial results from its application to young stars within stellar associations in sectors 1-5 of the \textit{TESS} data. While no new exoplanet candidate signals were found in this work, interesting results included the recovery of all known 2min TOIs around young stars in sectors 1-5 from 30min data alone, the recovery of the young exoplanet DS Tuc Ab, a number of young eclipsing binaries and a wide array of interesting rotation. A sensitivity analysis was also undertaken for each star, showing how recovery of injected planets varied with both depth and period for each individual target. Challenges for future searches for young exoplanets are discussed, the largest being stellar rotation with periods less than 1 day and a lack of a large sample of confirmed young stars.
\end{abstract}

\begin{keywords}
planets and satellites: detection -- planets and satellites: general -- stars: rotation -- techniques: photometric
\end{keywords}

%%%%%%%%%%%%%%%%%%%%%%%%%%%%%%%%%%%%%%%%%%%%%%%%%%

%%%%%%%%%%%%%%%%% BODY OF PAPER %%%%%%%%%%%%%%%%%%

\section{Introduction}

Despite the wealth of over 4000 verified exoplanets known today\footnote{https://exoplanetarchive.ipac.caltech.edu (30 April 2020)} there are still many unanswered questions concerning the formation mechanisms and system evolution which has led to the observed distribution of these objects. In particular, planetary migration, dynamical interactions with nearby stars and stellar evolution all can have major effects on the distribution and architecture of the final exoplanet population. This is especially important early in a planet's life  (e.g. <1Gyr), where phenomena such as accretion \citep{Marley2007OnJupiters,Manara2019ConstrainingRates}, ionising radiation causing atmospheric loss \citep{Baraffe2003Evolutionary209458,Owen2019AtmosphericExoplanets} and dynamical interactions with other forming planetary system bodies \citep{Ida2010TowardStars,Schlichting2015AtmosphericImpacts} can significantly change the mass, radius and orbital parameters of early exoplanets. While traditional exoplanet studies have been biased towards older exoplanets due to their relatively quiet host stars, there is a strong case to be made for the search for younger exoplanets, where planets are undergoing the majority of their evolution \citep{Adams2006LongTermInteractions,Ida2010TowardStars,Spiegel2012SpectralScenarios,Ida2013TowardGiants}. Discovering exoplanets <1 Gyr old will thus help to probe the main causes of planetary evolution, and help to fill a key gap in our knowledge of exoplanet history. 

However, the host stars of young exoplanets provide significant challenges for discovery due to their typically increased activity, rotation rates and relative proximity to neighbouring stars \citep[e.g.][]{Koeltzsch2009VariabilityRegion,Sergison2015UntanglingAnalysis,Rivilla2015Short-Cloud,Mascareno2016MagneticSeries,Briceno2019ThePopulations,Cegla2019StellarDiagnostics}. The combination of these factors can result in periodic stellar variability that can be on the order of planetary signals in both period and intensity \citep{Armstrong2015K20,Cody2018AMissions}.

Despite these challenges a small number of young exoplanets have been found. In line with early exoplanet discoveries, the first exoplanets around young stars were found using the radial velocity method, including four Hot Jupiters found in the Hyades \citep{Sato2007ATauri,Quinn2014HDCLUSTER} and Praesepe \citep{Quinn2012TWOCLUSTER} open clusters. However, for planets less massive than Hot Jupiters, the radial velocity method was found to be severely hampered by the inherent stellar variability and radial velocity jitter of the young host stars \citep{Saar1997Activity-RelatedStars,Paulson2004SearchingActivity,Brems2019Radial-velocityAge}. Recent methods such as those employed by \citet{Korhonen2015StellarCycles,Rajpaul2015AData} have pushed this limit down to approximately Neptune-sized planets for some solar-like stars, however Earth-sized planets are currently still beyond the capabilities of these methods for young active stars.

The high photometric precision offered by the launch of the \textit{Kepler} satellite proved crucial in increasing this sample of young exoplanets, especially when deliberately pointed at young open clusters in the \textit{K2} extended mission. Most discoveries in this era were made by the \textit{Zodiacal Exoplanets In Time (ZEIT)} team, who found 17 planets between Feb 2016 and Oct 2018 \citep[e.g.][]{Mann2016ZodiacalCluster,Rizzuto2017ZodiacalK2,Rizzuto2018Zodiacal16}.
Other interesting discoveries of young exoplanets in K2 have included K2-33b - a Neptune-sized planet in the 5-10 Myr Upper Scorpius stellar association \citep{David2016AStar}, K2-136A c in the Hyades - the first Neptune-sized planet orbiting a binary system in an open-cluster \citep{Ciardi2018K2-136:Planet}, and the closely-packed system of four planets around a $\sim$23 Myr pre-main sequence star in the Taurus-Auriga star forming region \citep{David2019ATaub,David2019FourTau}.
The launch of the \textit{TESS} is now ushering in a new era of exoplanet discovery, including the discovery of two new young exoplanets: DS Tuc Ab in the 45 Myr Tucana-Horologium association \citep{Benatti2019AA,Newton2019TESSAssociation}, and the ~24 Myr AU Mic b in the Beta Pictoris Moving Group \citep{Plavchan2020AMicroscopii}. These recent new discoveries highlight the fact that it is now possible to find planets across a wide range of the early evolution of these systems, but considerably more planets are needed at these ages before reliable statistics can be generated. %An ideal goal for this search is thus a timeline of planets with well-constrained ages, which can be used to inform the understanding of early planet formation and evolution.

A particularly promising place to look for these young exoplanets is within stellar associations. These are groups of gravitationally unbound stars which have a common origin, so still retain a common proper motion across the sky. Their relatively diffuse nature compared to similarly young star clusters makes them far more suited to study with \textit{TESS} given the relatively large pixels of the satellite \citep[angular resolution $\sim$21",][]{Vanderspek2018TESSHandbook}. Stars in these associations share similar ages, positions, compositions and kinematics, meaning that precise stellar properties can be determined \citep{Torres2006AstronomyMethod}. This in turn provides a significant advantage for precise determination of exoplanet characteristics. Furthermore, the ages of stars within these associations are typically very well constrained, which will allow a more detailed timeline of planetary evolution to be assembled. The most extensive census of 'bona-fide' members of these stellar associations was assembled by \citet{Gagne2018BANYAN150pc} in the development of their BANYAN $\Sigma$ Bayesian membership tool, but has recently been further expanded by works such as \citet{Gagne2018BANYAN.Data} and \cite{Esplin2019AMasses} thanks to the release of Gaia DR2 \citep{GaiaCollaboration2018GaiaProperties}. The extremely precise astrometric data from \textit{Gaia} will doubtless allow for further expansions to the membership lists of these associations as the satellite's mission continues. Note that \citet{Bouma2019Cluster7} recently assembled a larger list of young stars in clusters and associations (of which the BANYAN sample is a subset) by considering a wide array of sources in the literature, however they admit that this sample was designed for "completeness, not accuracy". For the initial design of the pipeline constructed in this work the homogeneous BANYAN selection criteria focused solely on stellar associations was preferred.

In order to discover any exoplanets within these associations, a key challenge is dissociating true transit signals from instrumental systematics or stellar 'noise' (activity or variability of the host star). While instrumental trends are commonly removed using techniques like cotrending basis vectors \citep{Thompson2016KeplerManual}, spacecraft pointing-based decorrelation \citep{Vanderburg2014AMission,Aigrain2016K2SC:Regression} or subtracting trends shared by simultaneously observed nearby stars \citep{Kovacs2005ASurveys,Kim2009DetrendingSurveys}, the problem of dissociating stellar noise from transit signals remains a challenging one. A large array of different methods have been developed in an attempt to solve this problem. These methods can be largely separated into three major approaches: sliding filters (e.g. \citet{Savitzky1964SmoothingProcedures.}), fitting splines/polynomials (e.g. \citet{Vanderburg2016PLANETARYMISSION}) and gaussian processes (e.g. \citet{Aigrain2015PreciseMission}; Gillen \& Aigrain, in prep).
An in-depth review and comparison of the most commonly used stellar activity detrending methods can be found in \citet{Hippke2019WotanPython}.
% Explain why GPs are great for one target at a time, but less so for rapid transit searches?
% Pre-whitening - Aigrain 2004

The earliest wide-field survey, HATNet \citep{Bakos2004Wide-FieldDetection}, used the Trend Fitting Algorithm \citep{Kovacs2005ASurveys} and External Parameter Decorrelation \citep{Bakos2010HAT-P-11b:Field} to detrend its light-curves, while WASP \citep{Pollacco2006TheCameras} shortly after preferred the  \texttt{sysrem} algorithm of \citet{Tamuz2005CorrectingCurves} - which removes common time- and position-dependent trends by taking into consideration the weighted average magnitude residuals for all stars - coupled with a boxcar-smoothing technique to handle some of the more variable stars observed \citep{CollierCameron2006ASearches}. More recent ground-based surveys such as KELT \citep{Pepper2007TheSurveys} and NGTS \citep{Wheatley2018TheNGTS} have developed these techniques further, and even introduced more advanced detrending methods such as gaussian processes as they target more active stars \citep[e.g.][]{Costes2020NGTS-8bHot-Jupiters,Gillen2020NGTS1}. 
However, the largest source of light-curves in the search for exoplanets has been space-based surveys, the volume and variety of which has required the development of more complex and robust detrending techniques.
The most prolific transiting exoplanet surveys thus far have been the \textit{Kepler} \citep{Borucki2010KeplerResults} and \textit{K2} \citep{Howell2014TheResults} missions and the ongoing \textit{TESS} mission \citep{Rickeretal.2014TheSatellite}, all of which share braodly similar detrending techniques. Overviews of the \textit{Kepler}, \textit{K2} and \textit{TESS} SPOC pipelines can be found in \citet{Jenkins2010OVERVIEWPIPELINE}, at https://keplerscience.arc.nasa.gov/k2-pipeline-release-notes.html and in \citet{Jenkins2016TheCenter} respectively.
%After the removal of instrumental and spacecraft systematics, general data cleaning and the removal of noise shared with other nearby stars in the Pixel Calibration (CAL) and Photometric Analysis (PA) steps, the resulting data is prepared for exoplanet searches in a Pre-Search Data Conditioning (PDC) step [SOURCE]. This includes...
% Overview of Main Kepler and TESS pipelines
%Also talk about other K2 pipelines?

While these state-of-the-art detrending methods have been extremely effective at discovering new exoplanets around older stars, they can still struggle to disassociate true transiting signals from the complex activity of young host stars. This was particularly well illustrated by \citet{Hippke2019WotanPython} in construction of their WOTAN tool, where  only  ~35$\%$ of 0.5$R_{Jup}$ exoplanet signals injected into a sample of 316 young stars were recovered (irrespective of the detrending method chosen), compared to an almost 100$\%$ recovery rate for similar planets around less noisy stars. Interestingly however, each of the methods tested recovered slightly different populations of the injected signals, so combining all of the tested methods increased the percentage of recovered planets to 43.8$\%$. This discrepancy was further highlighted by \citet{Rodenbeck2018RevisitingB}, who showed that different detrending methods resulted in quite different conclusions when considering the potential super-moon around \textit{Kepler-1625 b}. It is thus clear that not only is there a need for new detrending methods specifically focused on young stars, but also that there is a distinct benefit of using multiple detrending methods on the same data-set, thus providing dual motivations for the construction of an alternative detrending method in this work. The development and current design of the resulting detrending pipeline used in this work is described in detail in section \ref{detrending}.

The remainder of this paper is organised as follows. Section \ref{Methods} discusses the young star target selection, observations and the choice of a FFI pipeline, before describing the methods used to clean additional systematics from the \textit{TESS} data, detrend stellar variability, search for transits and inject model transits for a sensitivity analysis. Section \ref{Results} then discusses important results seen as a result of the developed detrending techniques, including the recovery of known young exoplanets, TOIs, and eclisping binaries, followed by an overview of the sorts of rotation and activity seen in this sample and the results of the conducted sensitivity analysis. The implications of these results for the future of exoplanetary searches around young stars are then discussed in detail in Section \ref{Discussion}, before a summary and conclusion in Section \ref{Conclusions}.

% n.b. for citing with an e.g.: \citep[e.g.][]{Author2012}.
%%%%%%%%%%%%%%%%%%%%%%%%%%%%%%%%%%%%%%%%%%%%%%%%%%%%%%%%%%%%%%%%%%%%%%%%%%%%%%%%%%%%%%%%%%%%%%%%%%%%%%

\section{Target Selection, Observations and Detrending Methods} \label{Methods}

\subsection{Target Selection}

\begin{figure*}
	\includegraphics[width = \textwidth]{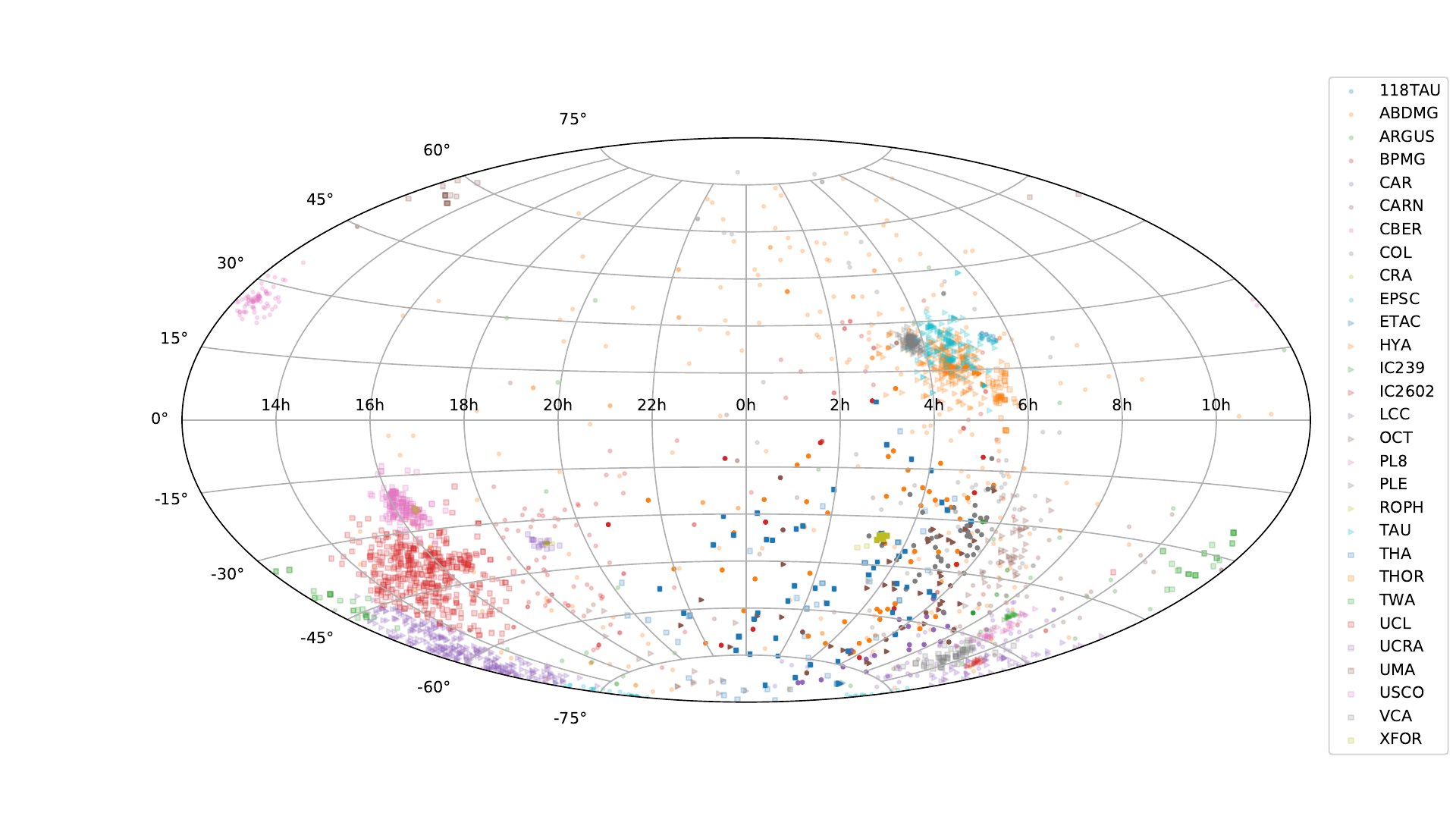}
    \caption{Equatorial sky distribution of initial young star sample, based on the bona-fide and high probability members of the 29 nearest stellar associations built into the BANYAN $\Sigma$ tool. The entire sample is shown in the background with the 256 sector 1-5 targets analysed in this work presented in solid colour. Abbreviations after those chosen in \citet{Gagne2018BANYAN150pc}.}
    \label{fig:sky_distribution}
\end{figure*}

While stellar association membership is now being expanded on a cluster by cluster basis thanks to the increased astrometric precision of the \textit{Gaia} satellite \citep{GaiaCollaboration2016TheMission,GaiaCollaboration2018GaiaProperties} (see for example \citet{Kuhn2019KinematicsDR2,Damiani2019StellarData,Zari2019StructureRegion}), a key list of 'bona-fide' stellar association members were assembled by \citet{Gagne2018BANYAN150pc} during preparation of the BANYAN $\Sigma$ Bayesian cluster membership tool. \citet{Gagne2018BANYAN150pc} consider a star to be a 'bona-fide' member if it has galactic XYZ UVW values consistent with those known for a given stellar association and exhibits an independent sign of youth. Such signs can include  mid-infrared excess, G-J vs GALEX NUV-G colours consistent with youth, X-ray emission with HR1 $\geq$ -0.15, lithium absorption above 100 mArmstrongs or a compatible luminosity class \citep{Gagne2018BANYAN.Data}. Combining  \citet{Gagne2018BANYAN150pc}'s initial census of bona-fide/high probability stellar association members with new high-probability members added in the two following BANYAN $\Sigma$ papers \citep{Gagne2018BANYAN.Data,Gagne2018BANYAN.2} yielded a total of 2977 objects spread over the 27 nearest known young stellar associations. \citet{Gagne2018VOLANS-CARINA:Pc} later expanded the BANYAN $\Sigma$ tool to also include the two new Argus \citep{zuckerman2019TheDisks} and Volans-Carina associations using the same membership criteria, so these were also added to the initial target list, to give the final distribution of 3076 young stars in stellar associations illustrated in Figure \ref{fig:sky_distribution}. For clarity an association by association breakdown of targets in the original BANYAN sample and the final sample of the sector 1-5 targets analysed in this work is presented in Table \ref{tab:Assos_breakdown}. This census remains the most extensive assemblage of bona-fide and high probability stellar association members with a common membership criterion, and hence represents a valuable starting place for the search for exoplanets around young stars in this work.

\begin{table}
	\centering
	\caption{Breakdown of association distribution for the complete BANYAN sample (centre) and final set of targets analysed in this work (right). The final targets consist those targets in Sectors 1-5 with available 30min-cadence light-curves from the \citet{Oelkers2018PrecisionApproach} pipeline, as discussed in Section \ref{pipeline choice}. Association abbreviations are based on those presented in \citet{Gagne2018BANYAN150pc} with the addition of ARGUS for the Argus association and VCA for the Volans-Carina Association.}
	\label{tab:Assos_breakdown}
	\begin{tabular}{ccc} % two columns, alignment for each
		\hline
		Association & BANYAN Targets & S1-S5 Analysed Targets\\
		\hline
		118TAU & 12  & 0\\
		ABDMG  & 298 & 55\\
		ARGUS  & 40  & 0\\
		BPMG   & 135 & 13\\
		CAR    & 85  & 19\\
		CARN   & 110 & 18\\
		CBER   & 76  & 0\\
		COL    & 107 & 43\\
		CRA    & 14  & 0\\
		EPSC   & 42  & 0\\
		ETAC   & 16  & 0\\
		HYA    & 241 & 7\\
		IC2391 & 14  & 0\\
		IC2602 & 16  & 0\\
		LCC    & 310 & 0\\
		OCT    & 101 & 40\\
		PL8    & 32  & 0\\
		PLE    & 204 & 0\\
		ROPH   & 182 & 0\\
		TAU    & 178 & 1\\
		THA    & 92  & 49 \\
		THOR   & 46  & 0\\
		TWA    & 50  & 0\\
		UCL    & 410 & 0\\
		UCRA   & 16  & 0\\
		UMA    & 13  & 0\\
		USCO   & 161 & 0\\
		VCA    & 59  & 0\\
		XFOR   & 16  & 11\\
		\hline
		Total  & 3076 & 256\\
		\hline
	\end{tabular}
\end{table}

The various BANYAN survey results were combined into a single tabular target list using the \textit{TOPCAT} table handling software \citep{Taylor2005TOPCATSoftware}. This list was then cross-matched with version 8 of the \textit{TESS} input catalog (TIC) \citep{Stassun2019TheList} using a 3 arcsecond radius. The Web \textit{TESS} Viewing Tool\footnote{https://heasarc.gsfc.nasa.gov/cgi-bin/tess/webtess/wtv.py} on the NASA \textit{TESS} website was used to determine which sector each target would be observed in and in turn compile target lists for each individual sector. Of the original 3076 objects, 1832 of them were forecast to be observed in \textit{TESS}'s first year of observations, with the breakdown of sources to be viewed in each Southern Hemisphere sector outlined in Table \ref{tab:BANYAN_year1}. For this work, only those viewed in Sectors 1 to 5 were considered.  

\begin{table}
	\centering
	\caption{Overview of BANYAN bona-fide members of stellar associations viewed in the first year of \textit{TESS} observations. Note that sources in the overlapping regions of the \textit{TESS} sectors, such as those in the continuous viewing zone, are counted multiple times.}
	\label{tab:BANYAN_year1}
	\begin{tabular}{cc} % two columns, alignment for each
		\hline
		Sector & Number of Sources observed\\
		\hline
		1 & 118\\
		2 & 130\\
		3 & 124\\
		4 & 186\\
		5 & 336\\
		6 & 258\\
		7 & 163\\
		8 & 166\\
		9 & 272\\
		10 & 399\\
		11 & 750\\
		12 & 396\\
		13 & 200\\
		\hline
	\end{tabular}
\end{table}

%When expanding the target selection to lower mass associated stars, 3D plots of galactic XYZ position and UVW velocity were found to be particularly instructive for discerning the best way to split the sky area into areas surrounding specific clusters.

\subsection{Observations}

\textit{TESS} is the successor to the highly successful \textit{Kepler} space telescope, which is designed to survey $>$85\% of the sky to search for planets transiting bright host stars \citep{Rickeretal.2014TheSatellite}. \textit{TESS} launched on 18 April 2018 and is now is into year two of its primary mission, having completed observations of the 13 Southern Hemisphere sectors. Sectors 1-5 considered in this work were observed between 25 July and 11 December 2018.

Two main data products are produced by the \textit{TESS} mission: 2min cadence light-curves (detrended and raw) and 30min cadence FFIs. 2min cadence light-curves are generated by Science Processing and Operations Center (SPOC) pipeline \citep{Jenkins2016TheCenter} for approximately 300,000 of the most promising stars (prioritized  by the smallest transiting planets that can be detected) from the \textit{TESS} Candidate Target List \citep[CTL,][]{Stassun2018TheList,Stassun2019TheList}, and are then made accessible to the public via the Mikulski Archive for Space Telescopes (MAST\footnote{https://mast.stsci.edu/portal/Mashup/Clients/Mast/Portal.html}). However, while these primary data products are very powerful for the main \textit{TESS} mission, because of the limited data transfer rates of the \textit{TESS} primary mission far from all of the stars in \textit{TESS}'s field of view will have 2min light-curves generated. Instead, light-curves from these stars can be retrieved from the 30min cadence Full-Frame Images. In addition, because of the increased activity and rotation rate of young stars discussed above, the standard detrending methods built for the 2min light-curves have difficulty flattening the light-curves for transit searches, and may even introduce confusing additional artefacts as a result of detrending \citep{Hippke2019WotanPython}. Because of the greater coverage offered by the 30min cadence data and concerns about detrending artefacts, the 30min cadence data were thus chosen for the initial transit search, with the simple aperture photometery (SAP) 2min cadence light-curves (where available) used as a secondary check.

An additional challenge common to both the 2min and 30min cadence data is the reasonably large pixels of the \textit{TESS} focal plane \citep{Vanderspek2018TESSHandbook}, which leads to significant blending of stars in crowded regions. This precludes the use of \textit{TESS} photometry alone for confirming transits in the dense centres of young clusters without confirmation from higher resolution instruments such as the \textit{SPITZER} Space Telescope \citep{Werner2004TheSpitzerTelescopeMission} or the cameras of the \textit{Next Generation Transit Survey - NGTS} \citep{Wheatley2013NextNGTS}. The power of \textit{NGTS}'s comparative resolution for identifying the true photometric source for \textit{TESS} objects is well-illustrated by \citet{Jackman2019NGTS-7Ab:Dwarfb} in the discovery of NGTS-7ab.

\subsection{Choice of FFI pipeline} \label{pipeline choice}

A number of methods currently exist for extracting light-curves from the \textit{TESS} FFIs. The simplest approach is to perform simple aperture photometry on the raw FFIs by overlaying an aperture over a cut-out around the object of interest, and summing up the flux under the target aperture for each cadence. This can be performed easily using the prebuilt \textit{lightkurve}\footnote{https://github.com/KeplerGO/lightkurve} \texttt{Python} package \citep{LightkurveCollaboration2018Lightkurve:Python}, and observing the resulting target-pixel-file provides an instructive look at the area around the star of interest. However, the simplistic nature of this anaylsis yields relatively noisy light-curves uncorrected for issues such as spacecraft pointing, jitter and localised scattered light. A greatly improved simple aperture photometry pipeline which accounts for these issues has been built in \texttt{Python} by \citet{Feinstein2019Eleanor:Images}. Christened \textit{eleanor}\footnote{https://github.com/afeinstein20/eleanor}, this package performs background subtraction, removal of spacecraft systematics such as jitter and pointing drift, and aperture/psf photometry. In addition, it provides tools to complete further systematics-removal via principal component analysis or psf-modelling. Light-curves from \textit{eleanor} will eventually be hosted on \textit{MAST}, but at the time of writing is only available as an open-source tool designed to work for \textit{TESS} sectors 1 and 2. 

An alternative difference imaging approach for 30min light-curve generation has been pursued by \citet{Oelkers2018PrecisionApproach}. Attempting to overcome the challenges posed to aperture photometry by \textit{TESS}'s large pixel sizes, this approach extracts light-curves via difference imaging analysis (DIA). In this method, one frame is blurred to the seeing conditions of the next before the two are subtracted from each other in order to retain only the variation in flux of stars between frames. Highlighting stars of interest within this process aids removal of contaminating stars in crowded regions and hence improves the light-curve extraction compared to standard aperture photometry methods. A similar technique has been used in ground-based surveys such as the Kilodegree Extremely Little Telescope \citep{Pepper2007TheSurveys,Siverd2012KELT-1b:Star}, though with a simpler Gaussian kernel than the Dirac $\delta$-function kernel used here. A full description of the DIA technique as applied to \textit{TESS} light-curves in sectors 1 and beyond can be found in \citet{Oelkers2018PrecisionApproach,Oelkers2019Beyond}. Light-curves extracted via this pathway are accessible from the Filtergraph data visualization service.\footnote{https://filtergraph.com/tess\_ffi}. It is further worth noting that as this document was being prepared \citet{Bouma2019Cluster7} released an alternative difference imaging pipeline for extraction of 30min FFI light-curve images which they applied to sectors 6 and 7, however these sectors are outside the scope of this work.

One further pipeline which can be used to extract 30-minute light-curves from the raw FFIs is the TASOC pipeline, under development by the \textit{TESS Asteroseismic Science Consortium} (\textit{TASC}). Heavily based on the K2P$^2$ pipeline \citep{Lund2015K2P2Mission}, this pipeline aims to supply \textit{TESS} photometry data for use in asteroseismology. Data from this source can be accessed online\footnote{http://tasoc.dk} after joining the consortium. So far raw light-curves extracted from the FFIs are available for sectors 1 and 2 \citep{Handberg2019TDA1+2}, while the open-source code used to extract these images from the FFIs can be found on github.\footnote{https://github.com/tasoc} However, one must be somewhat careful when using this source for transit-searches, as the primary goal of this group is asteroseismology, so at times light-curves may have been extracted in a way that prioritises the variability of the stellar signal over potential transit-like events. While this pipeline is still currently under development, a useful overview of its design and aims has been written by \citet{Lund2017DataTESS}.

Given the increased availability and comparatively clean light-curves provided by the DIA pipeline of Oelkers and Stassun  \citep{Oelkers2019Beyond,Oelkers2018PrecisionApproach}, this pipeline was chosen to extract light-curves from the \textit{TESS} FFIs in this work, resulting in light-curves for 256 individual objects. 
%This light-curve was then transformed into a \textit{LightCurve} class object in from the open source  \textit{lightkurve}\footnote{https://github.com/KeplerGO/lightkurve} Python package for ease of light-curve manipulation.
%However, for stars not available through this pipeline (usually because they fell below a contamination ratio of <5), the \textit{eleanor} pipeline was used, based on the eleanor pca flux.

\subsection{Removal of additional systematics from 30min light-curves}

Unlike the 2min PDCSAP light-curves retrieved from MAST, the 30min light-curves supplied by \citet{Oelkers2019Beyond} have not undergone the in-depth quality analysis completed by the SPOC pipeline, and as such still include some less-trustworthy epochs of increased pointing jitter, regular spacecraft momentum dumps and known data anomalies. It was thus necessary to remove these systematics before activity detrending and transit searches could begin.

The first step undertaken was to cut any epochs where fine pointing was known to have been lost, or other spacecraft anomalies affected the data. This was achieved by consulting the \textit{TESS} data release notes\footnote{Available at https://archive.stsci.edu/tess/tess\_drn.html} for each sector. This primarily affected sectors 1, 3 and 4. In Sector 1 a period of anomalously high pointing jitter was seen between approximately TJD 1347-1349 due to problems with the fine-pointing calibration. This was observed to be particularly bad between TJD 1348-1349.29, so all epochs between these times were masked from the analysis. Similarly in sector 3 a few experiments on the attitude control system (ACS) were undertaken by the \textit{TESS} team, dramatically increasing the scatter at these times. As a result, only data between TJD 1385.8966-1395.4800 in orbit 13 and TJD 1396.6050-1406.2925 in orbit 14 are scientifically useful, and all data at other epochs in this sector was cut. Sector 4 on the other hand was plagued by an instrument anomaly between TJD 1418.54 and TJD 1421.21, where communication was lost between the instrument and satellite. As a result, no data or telemetry was collected for this period, and some systematic trends were introduced following activation of the on-board heaters. Additional strong glints between TJD 1422.2297 - 1423.5020 (orbit 15) and TJD 1436.1047 - 1436.8353 (orbit 16) also plague some of the light-curves in sector 4 (particularly those on camera 4), however the amplitude and duration of these appear to vary between different targets, so may be better examined on a case by case basis.

\begin{figure}
	\includegraphics[width =\columnwidth]{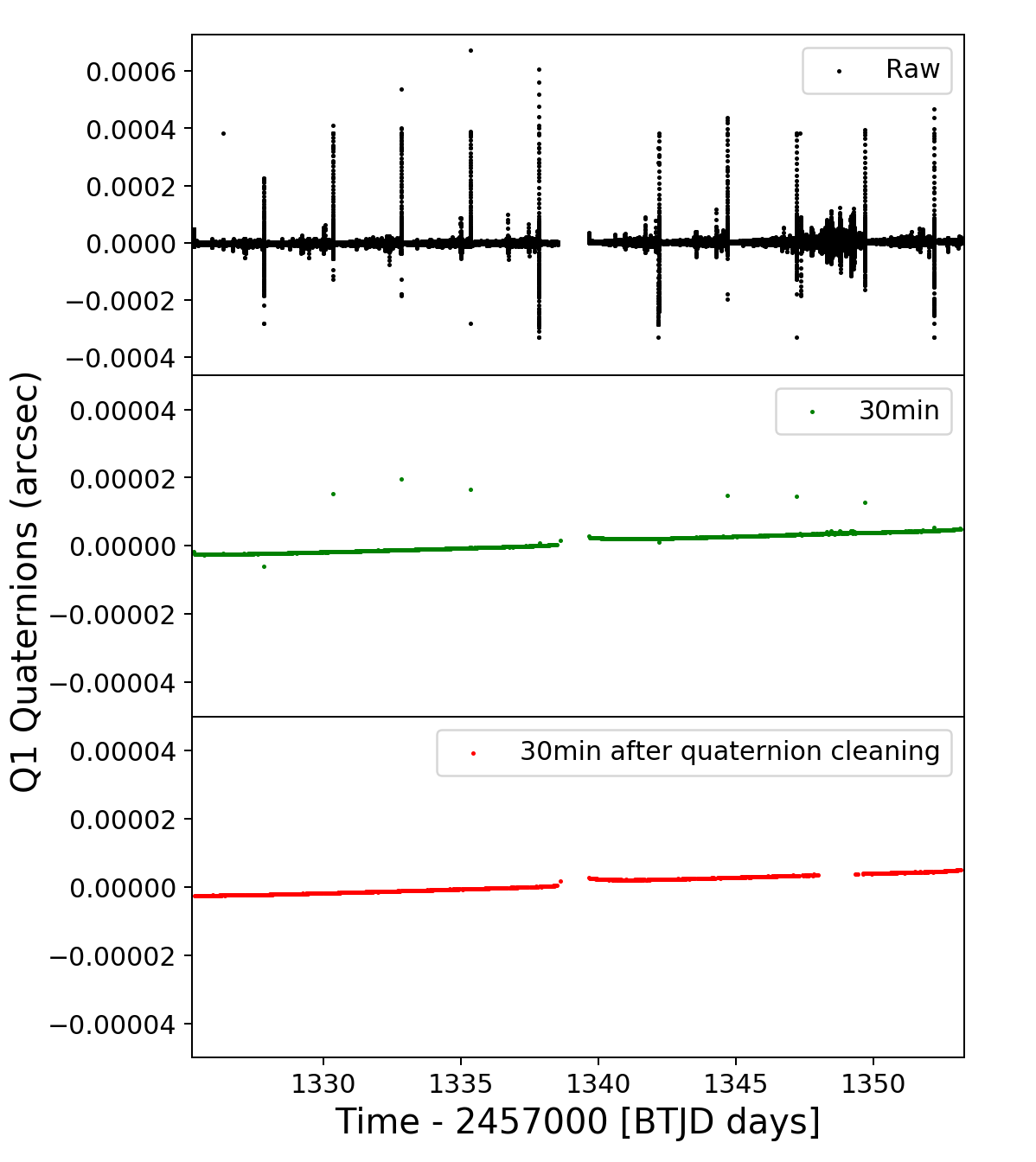}
    \caption{Q1 Engineering quaternion data for Sector 1, Camera 1 showing a clear increase in scatter around the 2.5d momentum dumps and the loss of fine pointing between TJD 1347-1349. Top: Raw 2s Engineering Quaternion data. Middle: Quaternion data binned into 30min bins to show typical number of affected data points. Bottom: Final quaternions after the removal of known periods of increased scatter and quaternion-based cleaning.}
    \label{fig:Quaternions_compilation}
\end{figure}

Following the removal of these sections of unreliable data, a more automated method was required to identify and remove additional epochs of increased spacecraft scatter on an epoch-by-epoch time-frame, such as those around the regular spacecraft momentum dumps. This was completed by generating scattering quality masks based on the engineering "quaternion" data, using a similar method to that described by \citet{Vanderburg2019TESS858} to prepare sector 3 30min cadence data in the discovery of multiple super-Earths around HR 858. The \textit{TESS} quaternion data\footnote{Available at https://archive.stsci.edu/missions/tess/engineering/} consists of 2s-cadence time-series data for each sector, describing attitude changes in three primary vectors (Q1, Q2, Q3) based on deviations from a selection of local guide stars. This provides a sector-specific overview of the spacecraft attitude and thus allows the generation of scatter-based quality masks for all targets in the \textit{TESS} aperture. An example of this data (both raw and binned into 30min bins) can be seen in Fig \ref{fig:Quaternions_compilation}.  To identify epochs with excessive scatter, the standard deviation for each vector was calculated, and any epoch with pointing scatter $\geq$5 standard deviations from the mean was flagged for removal. By combining the results from all three vectors into a single mask, all epochs within 0.01d of these points were removed from the data-set. This step efficiently removed all spurious signals relating to momentum dumps, as well as any remaining short periods of overly large scatter that were not picked up in the initial wider cuts.

For sources of particular interest which also possess \textit{TESS} 2min quality flags, it may also be advisable to implement the quality flags from the SPOC pipeline to the 30min data (perhaps in a similar manner to the conservative approach taken by \citet{Bouma2019Cluster7} in building their difference-imaging extraction pipeline for sectors 6 and 7) in order to remove the effects of additional systematics such as cosmic rays. However, given the lack of 2min light-curve availability for many of the objects in this sample, this feature was not implemented in the development of this initial pipeline. 

Following the removal of such sector-specific effects, the data was split into separate sections wherever there was a data gap of more than 0.1 days (to reduce the effect of flux jumps often seen after data gaps), except where doing so would result in fewer data points than the prescribed detrending window length discussed below. 

\subsection{Detrending of Stellar Variability} \label{detrending}

\subsubsection{Choice of base-detrending method}

Given the challenging range of activity-related and intrinsic stellar variability seen in the light-curves of the young star sample (and the associated difficulty of modelling so many individual light-curves in an initial transit search), it was considered wise to approach the detrending problem from the ground up, rather than necessarily relying on more traditional methods such as the Savitzky-Golay filter \citep{Savitzky1964SmoothingProcedures.}. A number of different detrending techniques were trialled early in the process, including simple low-order polynomial fitting, sinusoidal modelling, Savitky-Golay filtering and general smoothing over a range of window sizes. Of these, low-order polynomial fitting and the smoothing methods proved most successful at recovering injected transit signals (see Section \ref{injected_transits} for more information on the injected transit method used). 

A search for a method to combine and improve both of these methods led to LOWESS smoothing, or Locally Weighted Scatterplot Smoothing. This method, developed originally by \citet{Cleveland1979RobustScatterplotsb}, is a local polynomial regression method which works by fitting a low-order polynomial to a subset of the data (the width of which is set by a user-defined window) at each point along the x-axis using weighted least-squares regression. Under the weighted least-squares regression method, points nearer to the data-point being estimated are given more weight than those further away in the window. This weighting is one of the key differences between this method and the more commonly used Savitzky-Golay filter \citep{Savitzky1964SmoothingProcedures.}, and is particularly important for this application given the often swift evolution of young star light-curves. Indeed following the choice of this LOWESS-smoothing method, it was independently highlighted to be one of the best-performing detrending methods for the young-star sample tested by \citet{Hippke2019WotanPython}. In exoplanet literature however no other mention of LOWESS smoothing for exoplanet searches was found, with even the related LOESS smoothing method appearing quite rarely, despite having been used in detrending the TRAPPIST-1 system \citep{Luger2017ATRAPPIST-1} and in the Autoregressive Planet Search of \citet{Caceres2019AutoRegressiveMethodology}.

For this work, the standard tricube weighting function ($w = (1-|x|^3)^3$) was used, and the number of residual-based re-weightings retained as the default value of 3. The delta parameter was retained as $delta = 0.0$ as the sector-by-sector datasets were not overly large, but this can be adjusted if further speed enhancements are desired. Through experimentation a window size of 30 FFI data points (15hrs) was found to yield a good compromise between preserving the shape of injected transits and smoothing the stellar activity and variability of the host stars, except in more rapidly evolving light-curves, where a window size of 20 data points (10hrs) was found to be more appropriate. 
%As such the 'frac' parameter (the fraction of data-points to fit a polynomial to at each step) was set as \texttt{frac = 30/length(data)} (or \texttt{frac = 20/length(data))}, which allowed for consistent window sizes even when there were gaps in the dataset. For stars where 2min light-curve data was available, these were proportionally increased to windows of 450 and 300 respectively.
%Maybe add in specific example of where 20-box lowess is better - e.g. how rapidly evolving??

\subsubsection{Removal of Peaks and Troughs} \label{peak_cutting}

\begin{figure}
	\includegraphics[width =\columnwidth]{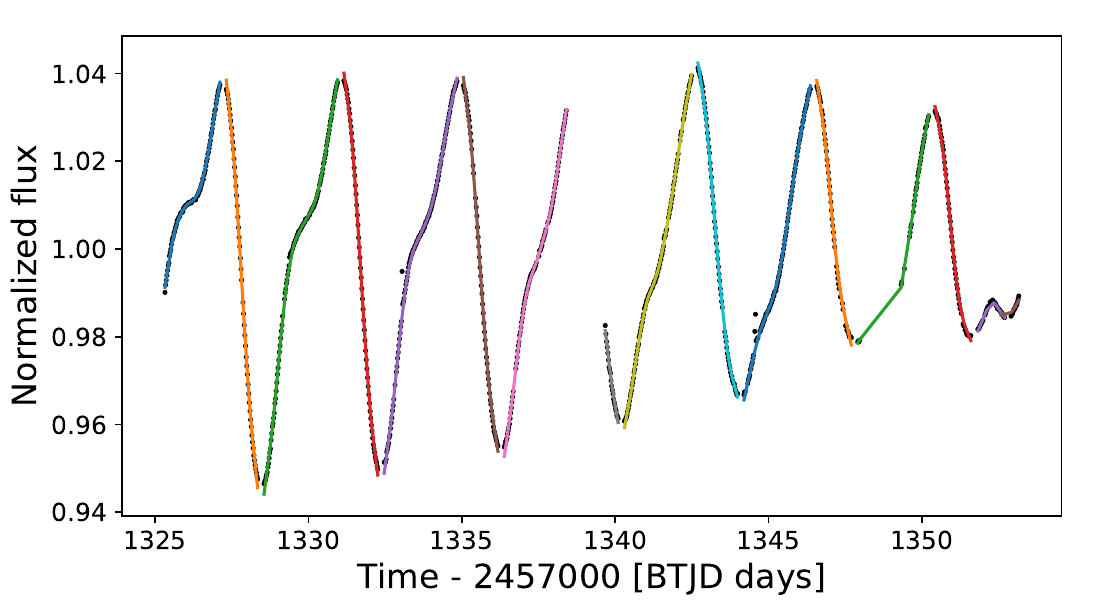}
    \caption{Example of the peak cutting technique in use on the rapidly evolving light-curve for HIP 32235. The applied 20-box partial LOWESS-smoothing model used for detrending the light-curve is also over-plotted in colour. Note that the change in colour between subsequent sections of the LOWESS-detrending shows where the light-curve has been split after any 0.1d gap in the data. This typically occurs after each peak or trough, or where significant detrimental scatter has been removed.}
    \label{fig:Peak_cut_example}
\end{figure}

One of the other key challenges present in using a Box Least Squares (BLS) search for light-curves from young active stars is that the troughs of regular stellar activity (e.g. from the rotation of star-spots) are often picked up as the largest peaks in the BLS periodograms, even after LOWESS detrending. This is particularly the case for rapidly rotating stars, or those with short-period intrinsic variability. Such sources typically exhibit sharp peaks and troughs in the extracted 30min light-curve, which is often highlighted in the BLS search. In an attempt to combat this problem for light-curves with particularly sharp stellar activity, the effect of cutting the peaks and troughs of this stellar activity was tested. To begin with, the peaks were located using the \texttt{find\_peaks} function from the \texttt{scipy.signal} library\footnote{https://docs.scipy.org/doc/scipy/reference/signal.html} \citep{Virtanen2019SciPyPython} using a required prominence of 0.001 and width of 15 data points (7.5hrs). This helped to ensure that only wider peaks generally associated with stellar activity/variability were flagged as peaks or troughs, but the settings may need adjusting for some more complex light-curves. The activity/variability troughs were located using a similar \texttt{find\_peaks} search on a negative version of the light-curve flux. Data points within 0.1d either side of each peak and trough were then cut from the light-curve before LOWESS-detrending was applied. An example of this in practice is shown in Fig \ref{fig:Peak_cut_example} for HIP 32235, a rotationally variable  G6V star in the Carina stellar association.

\begin{figure}
	\includegraphics[width =\columnwidth]{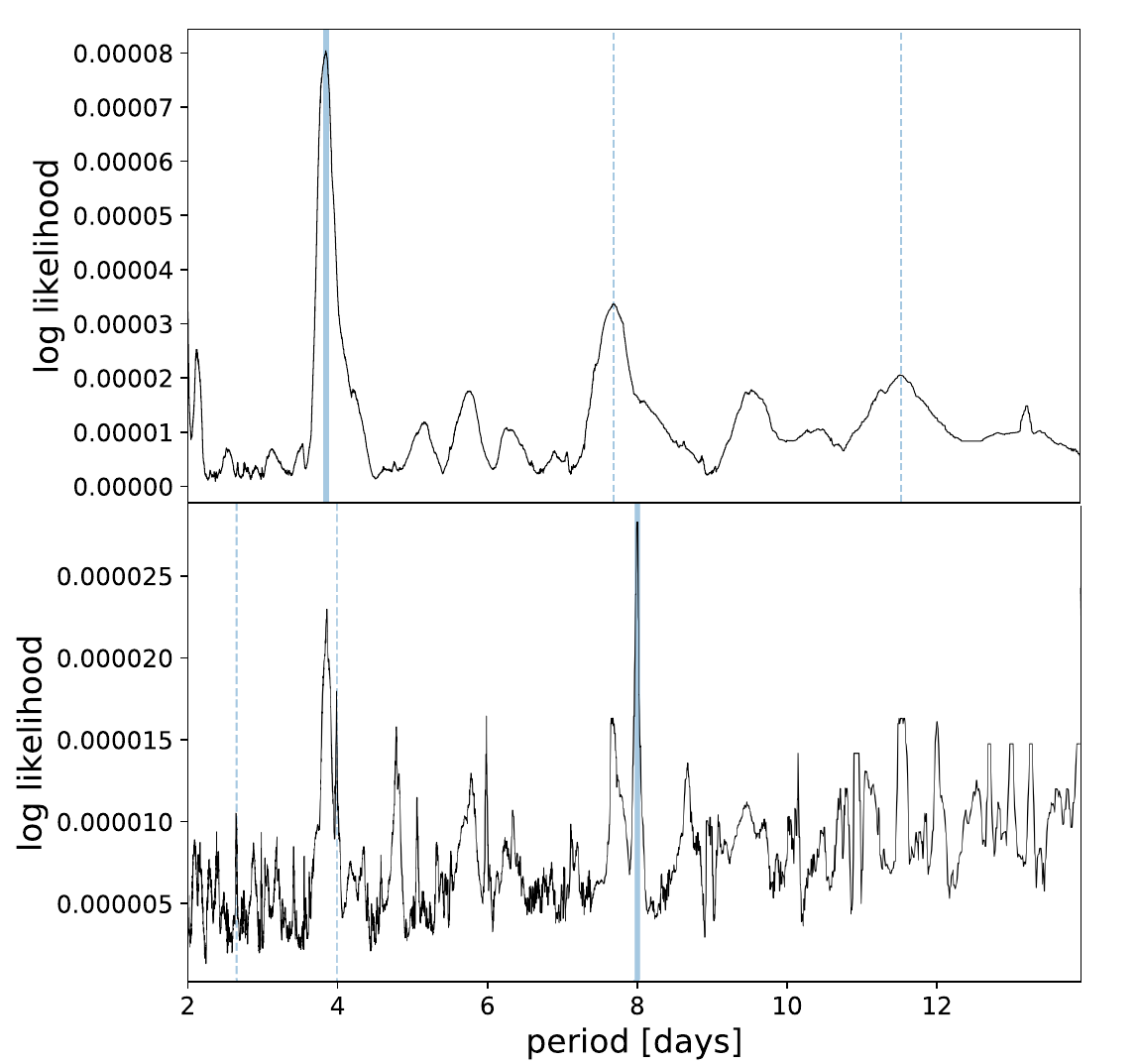}
    \caption{BLS periodograms for HIP 32235 following 20-bin LOWESS detrending without (top) and with (bottom) peak cutting  implemented. In this case peak cutting displays a clear benefit to the recovery of an injected $0.04 R_P/R_*$ 8d planet in the presence of rapidly evolving (3.84d period) stellar activity. The highest power period is highlighted in blue, with aliases of the same period shown by dotted blue lines.}
    \label{fig:Peak_cut_periodograms}
\end{figure}

The power of this peak-cutting method is illustrated for the same source in Fig \ref{fig:Peak_cut_periodograms}. When a 0.04 $R_P/R_*$ radius ratio planet is injected into the light-curve (using the method described in Section \ref{injected_transits}) and the light-curve is detrended using the LOWESS-based method described above (boxsize = 20), any signal of the planet is clearly overwhelmed by the 3.84d period rotational variability of the star, as is shown in the top of Fig \ref{fig:Peak_cut_periodograms}. However, applying the described peak-cutting technique to the data easily recovers the signal of the injected 8-day 0.04 $R_P/R_*$ planet, as illustrated in the lower half of Fig \ref{fig:Peak_cut_periodograms}. This technique is thus another powerful tool for pushing down to lower radii in the search for exoplanets around young active stars. In general, this peak-cutting method was found particularly effective for light-curves with sharp oscillations/rotations, or ones which were strongly periodic. However, it is important to note that this technique did not always offer improvements over the non-peak-cut method. A further discussion of when this technique is most effective can be found in section \ref{Peak_cutting_comparison}.

 %Indeed for many light curves where light-curve variability was observed to be less sharp (for example the clear sinusoidal variation observed for HIP 1113 - bottom panel, Fig \ref{fig:sinusoidal_var}) or less strictly periodic (e.g HIP 105388 - top panel, Fig \ref{fig:HIP_105388_AB_Pic}), the removal of such data near peaks and troughs was observed to decrease the amplitude of the recovered signal in the periodogram. %We thus offer this technique as a helpful extra tool in the light-curve detrending toolbox, but not one to be applied blindly without also trying a non peak-cutting method alongside it.

\subsubsection{Transit masking and light-curve interpolation} \label{transit masking}

\begin{figure}
	\includegraphics[width =\columnwidth]{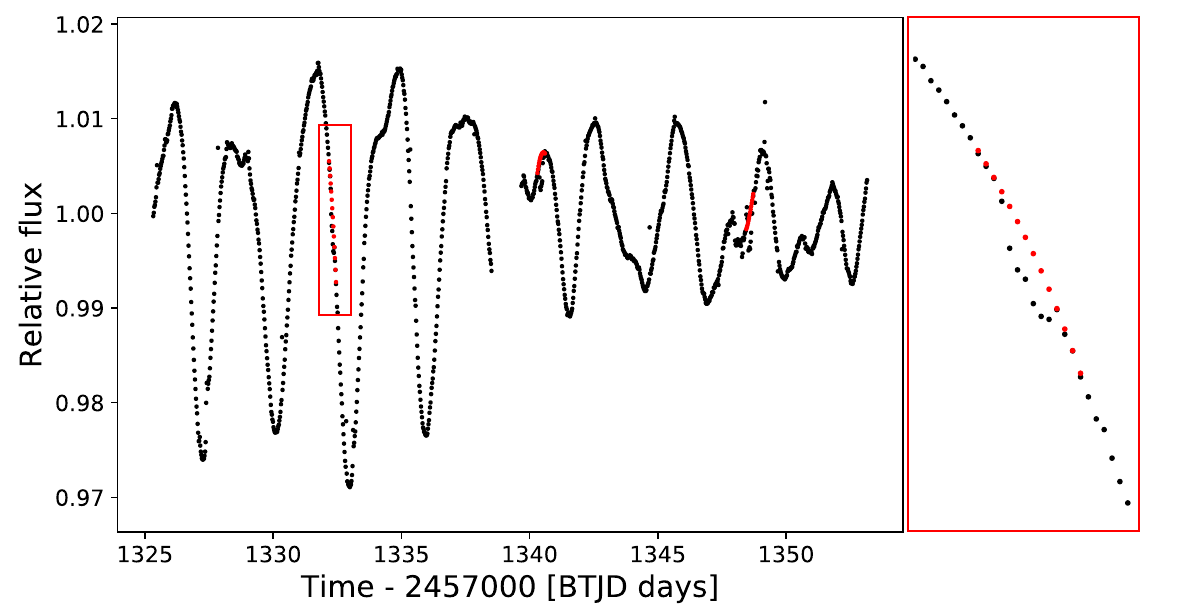}
    \caption{Example of interpolation over transit masked areas of the light-curve for DS Tuc A in order to best preserve the stellar activity signature through these segments. Original data points are shown in black, while interpolation over the removed transits is shown in red. The red box on the right shows detail of the area in the vicinity of the first transit.}
    \label{fig:Transit_mask}
\end{figure}

Transit masking is a commonly used method in planetary and eclipsing binary science to independently detrend long-term variability in stellar light-curves without changing the shape of the transit curve, and is implemented into most common light-curve manipulation tools (e.g. \citet{Luger2016CURVES,LightkurveCollaboration2018Lightkurve:Python}). However, due to the often rapid evolution of young star light-curves, simply cutting out the data near a suspected transit before detrending can lead to spurious variability signals over the duration of the transit. In order to combat this issue in the developed detrending method a two-step transit masking approach was used which accounts for the brightness variation of the host star during the transit. Firstly, a mask is generated based on a period, epoch and duration for the suspected transit (either user-defined or from a previous BLS search), and the new light-curve generated by removing any points within the selected transit duration. Then, in order to account for the stellar flux variability over the course of the transit, the new holes in this light-curve are refilled using a quadratic interpolation (using \texttt{scipy.interpolate.interp1d}) between the cut points. It is this new flux array featuring interpolated sections over the transit mask that is then used in the LOWESS-detrending step, and in turn divided out to form the final detrended light-curve. The importance of using such a method is illustrated using the source DS Tuc A in Fig \ref{fig:Transit_mask}. It is immediately obvious that not taking into consideration the variation in stellar flux around the epoch of the first cut transit (approx TJD 1132) would lead to a considerably different transit shape after the main LOWESS-based detrending was is undertaken. A full discussion of the full detrending of DS Tuc A and the resulting recovery of the young exoplanet DS Tuc Ab can be found below in Section \ref{Recovery of DS Tuc A b}. 

The authors make available a basic version of this detrending code online. \footnote{https://github.com/mbattley/YSD}

\subsection{Transit searching algorithm} \label{bls}

The standard Box Least Squares method was used to search for periodic transit-like signals in the detrended light-curves. This method, originally developed by \citet{Kovacs2002ATransits}, fits a series of box-shaped dips at a range of periods in order to generate a periodogram comparing the relative strengths of the different period hypotheses. The specific implementation used in this code is the \texttt{astropy.timeseries.BoxLeastSquares} method\footnote{https://docs.astropy.org/en/stable/api/astropy.timeseries.BoxLeastSquares.html}, which is a \texttt{python} implementation of the computational method described by \citet{Hartman2016VARTOOLS:Data}. In this work the strongest BLS peak and the two next strongest non-harmonic peaks were investigated.
%In this method one must supply a grid of durations (set between 0.05-1d in this case), a mininmum number of transits (typically three here, but also tested for two in sector 3 given the much shorter duration of useable data in this sector) and a frequency factor to trade off efficiency with period precision (this was set equal to 1.0 initially, but was adjusted to 5.0 for sources of particular promise).

% Chose BoxLeastSquares over Transit Least Squares because of 30min cadence having less info about ingress and egress

\subsection{Injected transits} \label{injected_transits}

In order to test the sensitivity of this new detrending method to finding planets around young stars, a series of model planet transits were injected into the light-curves before detrending. For this analysis eight different orbital periods were tested (1.0, 2.0, 4.0, 6.0, 8.0, 10.0, 12.0 and 14.0 days), with the epoch chosen randomly for each injection. Note that unlike ground-based surveys, \textit{TESS} does not exhibit 1-day systematic errors due to the Earth's rotation which could adversely affect these period choices. Rather than injecting specific planet sizes, five planet to star ratios $R_p/R_*$ were tested in this sensitivity analysis: 0.1, 0.075, 0.05, 0.04 and 0.03, beyond which recovery was observed to be quite rare. For reference, around a sun-like star a radius ratio of 0.1 corresponds to an approximately Jupiter-sized planet and $R_p/R_*$ = 0.03 corresponds to a sub-Neptune-sized planet. Stellar parameters for each star were retrieved from TIC v8 \citep{Stassun2019TheList}. Where possible, orbital separation for each planet was then derived from Kepler's Third Law. If information on a star's mass or radius was not available, the corresponding planet was assigned an orbital separation of 17.0$R_*$, representing the average orbital separation for planets with 8d periods (in the middle of the period range) on the NASA Exoplanet Archive.\footnote{https://exoplanetarchive.ipac.caltech.edu/} Planet transits were generated using \citet{Kreidberg2015Batman:Python}'s \texttt{batman} python implementation of \citet{Mandel2002AnalyticSearches}'s transit model, assuming non-linear limb-darkening with coefficients [0.5, 0.1, 0.1, -0.1]. After injection, each light-curve was detrended using a standard 30-bin run of the LOWESS-smoothing pipeline (without peak-cutting implemented). Injected signals were considered to have been 'recovered' if they appeared as one of the three highest peaks in the BLS periodogram, ignoring harmonics of the maximum peak.

%The generated planet transits were injected into light-curves retrieved only from the \citet{Oelkers2018PrecisionApproach,Oelkers2019Beyond} pipeline for consistency. This constituted 256 individual sources in sectors 1-5, though many were observed over multiple sectors.

%%%%%%%%%%%%%%%%%%%%%%%%%%%%%%%%%%%%%%%%%%%%%%%%%%%%%%%%%%%%%%%%%%%%%%%%%%%%%%%%%%%%%%%%%%%%%%%%%%%%%%

\section{Results from sectors 1-5} \label{Results}

Despite 30min DIA light-curves only being available for 256 of the BANYAN objects within the first five sectors of \textit{TESS} data, a wide range of interesting activity was observed, from the recovery of known young exoplanets and \textit{TESS} objects of interest (TOIs) to an eccentric eclipsing binary system and a large variety of unusual rotation and activity profiles. Even this relatively small sample clearly demonstrates the unusually large variation in light-curves of young stars compared to their older counterparts, and consequently helps to explain why far fewer planets have thus far been found around stars of these ages. The conducted sensitivity analysis goes one step further, investigating the comparative recovery rates for different combinations of injected period and planetary radius. Some of the most interesting results are summarised below in this section.

\subsection{Recovery of confirmed young exoplanets} \label{Recovery of DS Tuc A b}

One of the most promising initial results from the application of this pipeline on the 30min data was the recovery of both of the known transiting exoplanets found around young stars in sectors 1-5: DS Tuc A b \citep{Benatti2019AA,Newton2019TESSAssociation} and AU Mic b \citep{Plavchan2020AMicroscopii}. The recovery of DS Tuc A b is described in detail here as an example of the full pipeline in use. 

DS Tuc A (TIC 410214986/TOI 200.01/HIP 116748 A) was observed in Sector 1 of the \textit{TESS} observations, carried out between 25th July - 22nd August 2018. It is a G6V type star known to be associated with the 45 Myr Tucana-Horologium association \citep{Zuckerman2000IdentificationFormation}. DS Tuc A fell on camera 3 of the instrument, and yielded approximately 27 days of photometry.

Interestingly, \citet{Newton2019TESSAssociation} admit in their work that the candidate signal was actually found by human eyeballing around a spurious activity-induced periodicity peak flagged by a run of the SPOC Transiting Planet Search (TPS) module on the 2min PDC-SAP data. However, a later archival TPS run of the SPOC pipeline (after the planet candidate was announced to the community as TOI 200.01) was seen to detect a periodic transit crossing event which passed all of \citet{Newton2019TESSAssociation}'s false positive tests. The planetary candidate was later confirmed using additional photometry, spectroscopic methods and high contrast imaging. 

\citet{Benatti2019AA} independently reprocessed the TESS data for this object using improved stellar parameters (including crucially accounting for dilution from DS Tuc B) and fitted two different models to determine planetary parameters. The first model involved modelling only the first two transits (on account of the large pointing jitter around the 3rd transit) using \texttt{PyOrbit} \citep{Malavolta2016TheCluster} to complete a simultaneous fit for modulation and a transit signal, along with \texttt{emcee} \citep{Foreman-Mackey2012Emcee:Hammer} and \texttt{PyDE} \citep{Storn1997DifferentialSpaces} to establish the most likely planetary parameters. Alongside this they tested modelling all three transits with the \texttt{batman} package \citep{Kreidberg2015Batman:Python} after applying a 0.55d 3rd order running polynomial to flatten the light-curve and analysed the posterior with \texttt{emcee}. While both methods yielded consistent results, they eventually adopted the planetary parameters from the first method and found a best-fit solution of a 0.5$R_{Jup}$ planet. \citet{Benatti2019AA} confirmed the planetary nature of this object using radial velocities from the HARPS spectrograph.

\begin{figure}
	\includegraphics[width = \columnwidth]{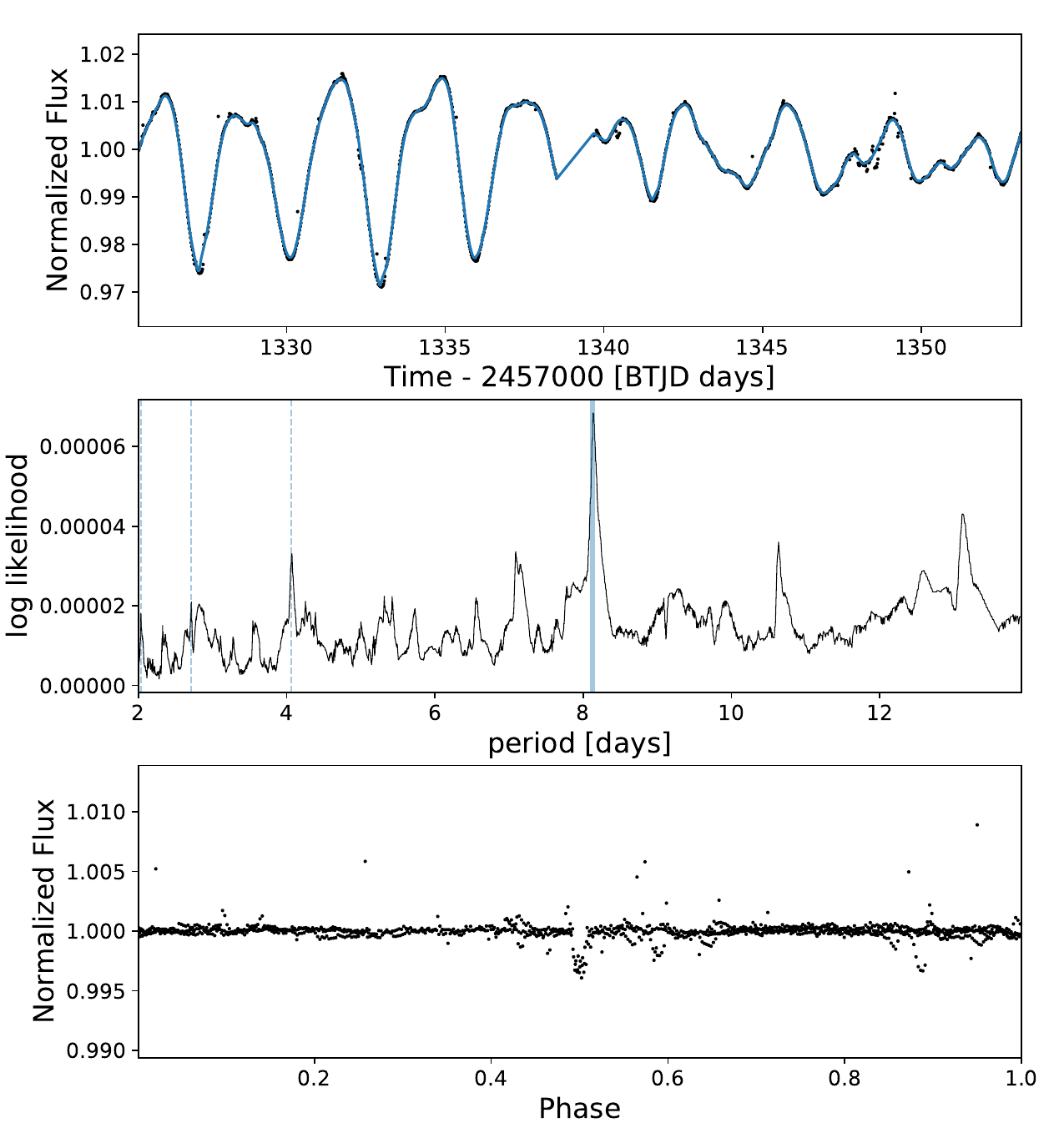}
    \caption{Example of the developed detrending pipeline in use, showing the recovery of the 45Myr exoplanet DS Tuc A b. Top: The original sector 1 light-curve for DS Tuc A with overplotted 20-bin LOWESS detrending in blue, including interpolation over suspected transits. Middle: BLS Periodogram for the the light-curve after detrending, with the peak period of 8.138d highlighted in blue (n.b. aliases of this period are shown by the blue dotted lines). Bottom: the resulting light-curve after the detrending pipeline has been applied, folded by the maximum peak of the BLS Periodogram}
    \label{fig:DS_Tuc_A_detrend}
\end{figure}

In this work, the planetary signal was highlighted as the highest peak of the BLS periodogram in a 20-bin LOWESS-partial run of the standard detrending pipeline, however the transits were also clearly visible by eye after the 30-bin LOWESS-partial run. Noting that the third transit fell in the period of heightened pointing jitter between TJD 1347-1349, this section of the data was unmasked for this target. The recovered period was 8.14d, in agreement with the accepted value from \citet{Newton2019TESSAssociation}. Using the period and epoch derived from the BLS search, the transit masking and light-curve interpolation technique described in section \ref{transit masking} was applied, completing the clear detrending and transit recovery of DS Tuc A b presented in Fig \ref{fig:DS_Tuc_A_detrend}. Note that in the bottom panel the light-curve has been folded by 8.138d, the accepted planet period from \citep{Newton2019TESSAssociation}. Interestingly, the use of the peak-cutting technique for this object was found to increase the significance of the true transit period in the 30-bin case (changing the 8.14d period-peak in the BLS periodogram from insignificant to the 6th strongest after peak-cutting), but decreased its significance for the 20-bin LOWESS-detrend.

% Once found, applied known P (8.138268d pm 1.1 x e-5), T_0 (1332.30997 pm 0.00026 TJD), duration = 0.18 (-0.12, +0.13) to tidy it up and the used a lowess 10-based smoothing (either full or partial work well)

The clear recovery of this Neptune to Saturn-sized planet using the 30min data alone bodes well for future exoplanet candidate discoveries from the Full Frame Images, thus demonstrating the wealth of knowledge to be gained from these images.

\begin{figure}
	\includegraphics[width = \columnwidth]{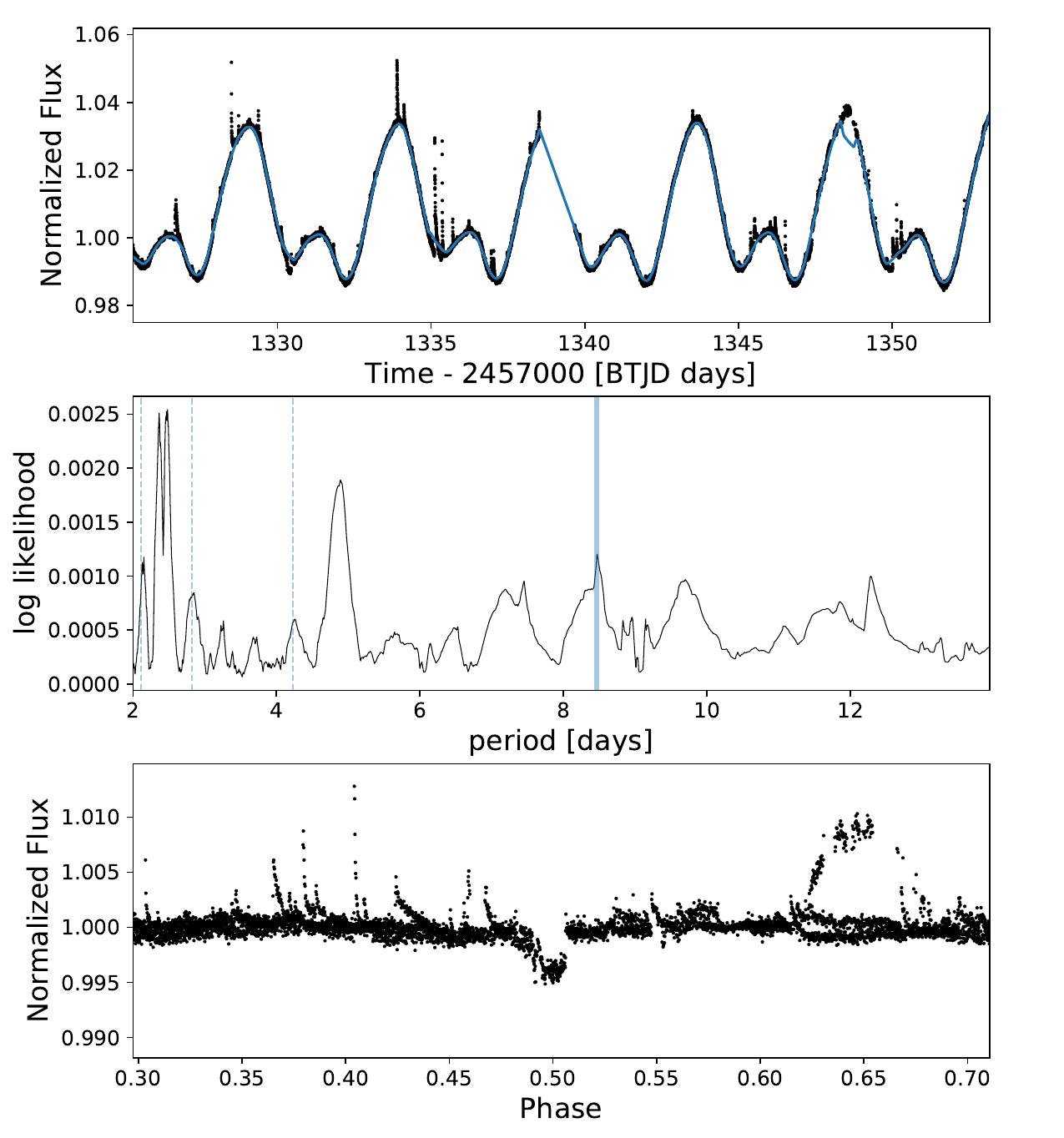}
    \caption{A 20bin-LOWESS smoothing run on the \textit{TESS} 2min cadence data for AU Mic, showing the recovery of the 8.46d planetary signal first identified by \citet{Plavchan2020AMicroscopii}. Format similar to that explained for Figure \ref{fig:DS_Tuc_A_detrend}, except that the final light-curve is folded by the third highest period, as highlighted in blue in the BLS periodogram.}
    \label{fig:AU_Mic_Detrend}
\end{figure}

% Add comment about AU Mic b?
Unfortunately data for AU Mic was not extracted by \citep{Oelkers2018PrecisionApproach}'s 30min pipeline, however application of this pipeline to the 2min data for AU Mic easily recovered the signal of the 8.46d planet proposed by \citet{Plavchan2020AMicroscopii}, as shown in Fig \ref{fig:AU_Mic_Detrend}. In this case a 20bin-lowess smoothing run revealed an alias 8.46d period as the third highest peak in the periodogram. In this particular case the peak-cutting option did not aid recovery of the planet substantially since the improvements of cutting the sharp troughs were balanced by the inadvertent cutting of the first transit. Nonetheless, the 8.5d signal remained the third highest peak after 20bin-LOWESS smoothing was applied to the peak-cut light-curve.

\subsection{Retrieval of other \textit{TESS} objects of interest}

A number of other interesting signals were independently found using this pipeline, including a pair of additional \textit{TESS} Objects of Interest (TOIs). In the interest of independence these signals were found with no prior knowledge of the TOIs in these sectors, but instead candidate signals highlighted in the BLS search in this work were cross-matched with the TOI list at a later date. This resulted in the recovery of two known TOIs: TOI 447.01 and TOI 450.01.

\begin{figure}
	\includegraphics[width =\columnwidth]{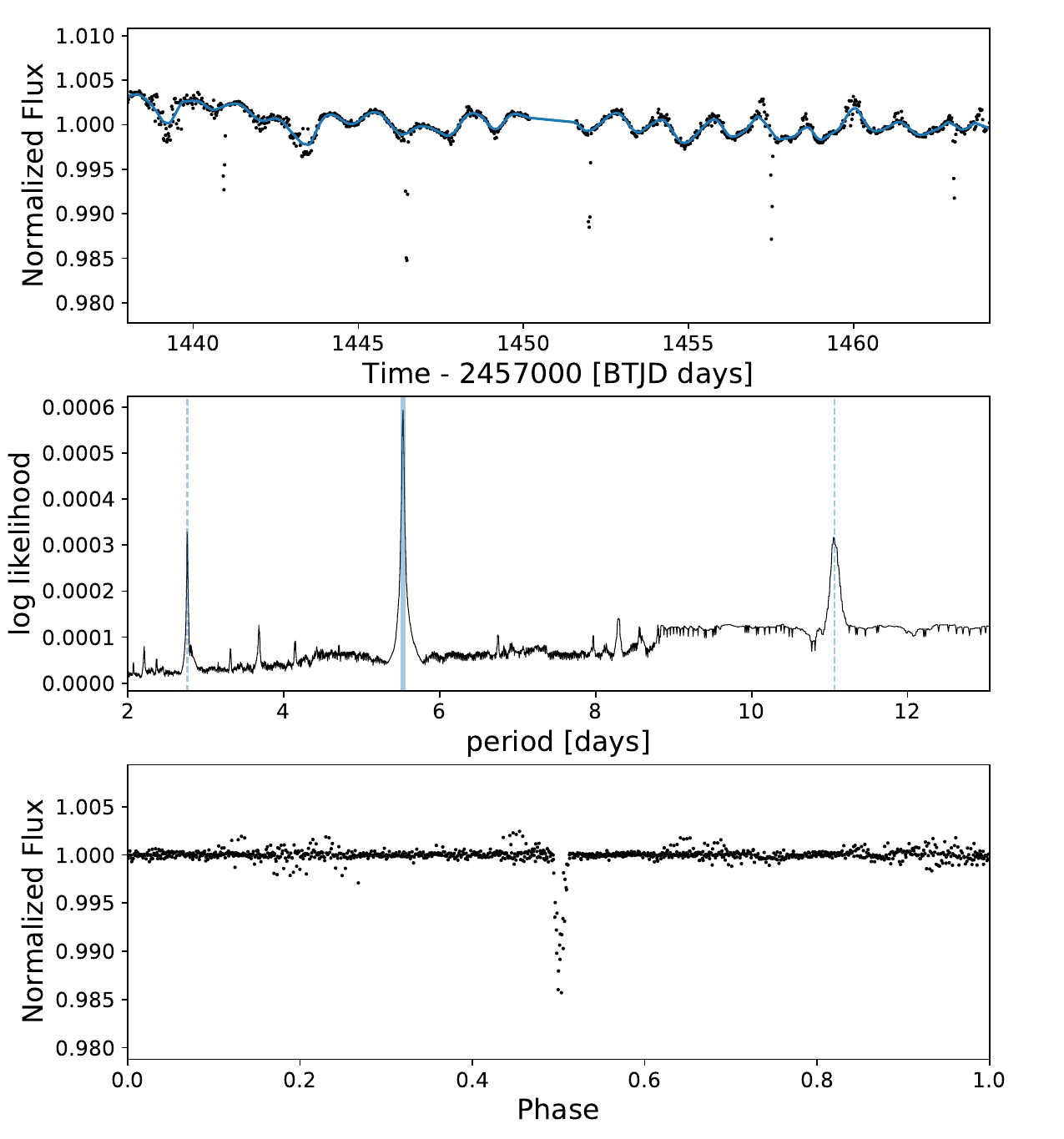}
    \caption{Recovery of TOI 447.01/HD 33512, a likely long-period binary in the 35Myr Octans Association. In this case the maximum period was found to be 5.528d. Format after that explained in Fig \ref{fig:DS_Tuc_A_detrend}.}
    \label{fig:Other_EB1}
\end{figure}

For TOI 447.01, a repeated dip of approximately 15mmag was highlighted by the BLS search in sector 5 with a period of 5.528 days around the F5/6V star HD 33512/TIC 14091633 (see Fig \ref{fig:Other_EB1}). This signal was uncovered using the a standard 30-bin LOWESS run of the pipeline, without peak-cutting. HD 33512 is a relatively bright (T=8.8) F5/6V star in the Octans association, giving it an approximate age of 35$\pm$5 Myr \citep{Gagne2018BANYAN150pc}, which would make this an interesting target for follow-up of young evolving systems if the planetary hypothesis was correct. As a TOI this target underwent follow-up by the TFOP team, and though initially seemed positive from the \textit{TESS} data alone, it was eventually flagged as a False Positive (FP) by the TFOP working group due to the observed linear drift in RV measurements, relatively large radius, and changing width of FWHM.\footnote{https://exofop.ipac.caltech.edu/tess/target.php?id=14091633} Nonetheless, this object is a likely long-period binary, so may still be an interesting system for the study of young binary systems.

\begin{figure}
	\includegraphics[width =\columnwidth]{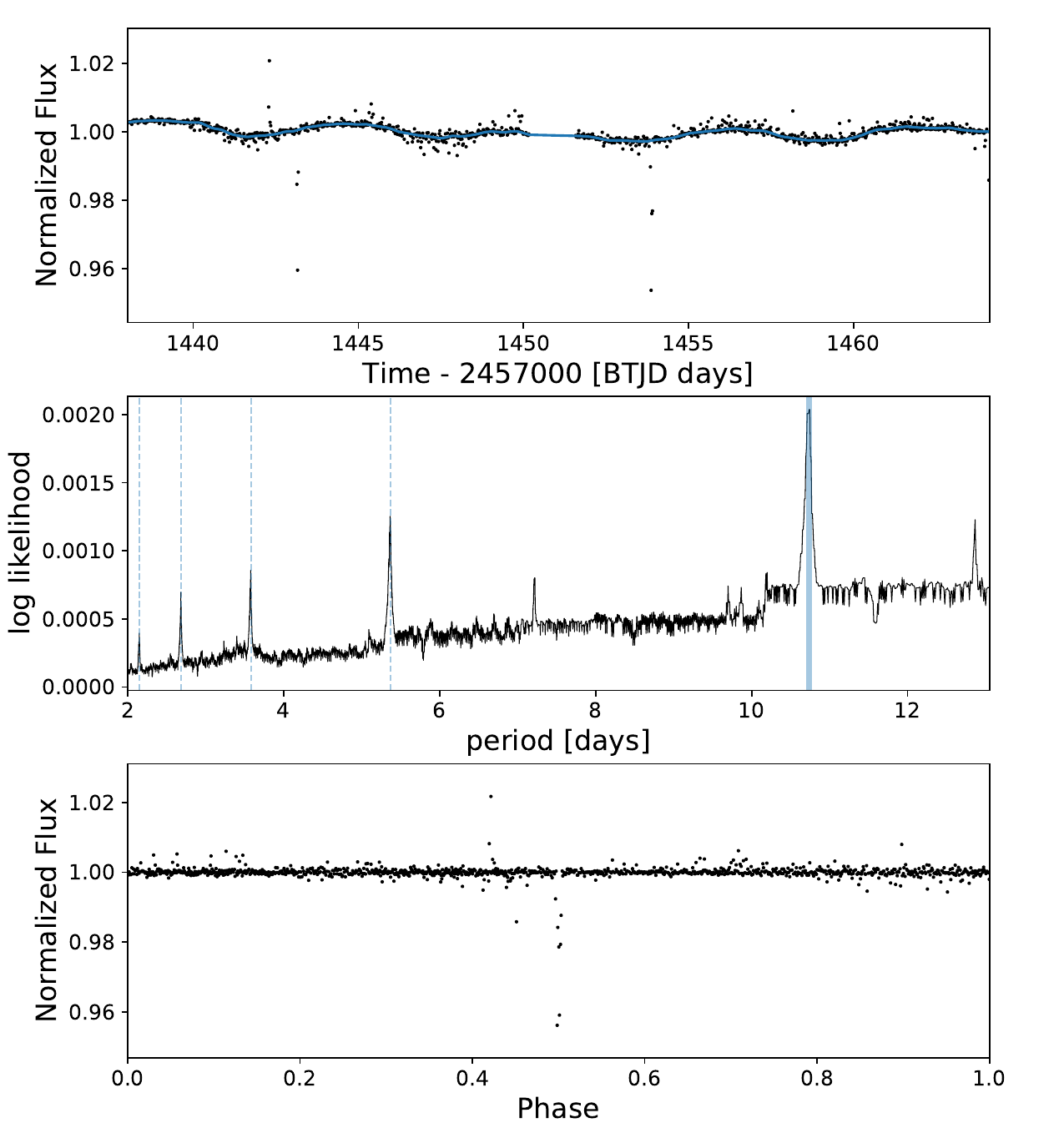}
    \caption{Recovery of TOI 450.01/J0516-3124, a spectroscopic binary in the 42 Myr Columba Association. In this case the maximum period was found to be 10.7d. Format after that explained in Fig \ref{fig:DS_Tuc_A_detrend}, except that overplotted LOWESS-detrending in the top panel has bins of 30 rather than 20.}
    \label{fig:Other_EB2}
\end{figure}

The second TOI independently recovered was TOI 450.01, around the source 2MASS J05160118-3124457/TIC 77951245. This source was observed in sectors 5 and 6 of \textit{TESS} observations, however alike to TOI 447.01 only sector 5 light-curves were available for it from the DIA 30min pipeline. In this work the candidate signal was initially revealed by a 30-bin LOWESS-partial run of the main detrending pipeline on sector 5. Two approximately 40mmag transits were observed in the flattened light-curve, with a recovered period of 10.74 days and initial epoch of 1443.2d, as shown in Fig \ref{fig:Other_EB2}. The parameters derived from this 30min data are slightly different to those derived from the 2min data by the SPOC TOI analysis (P = 10.7148$\pm$0.0001, depth = 63.6$\pm$0.8, epoch = 2458443.1686 $\pm$ 0.0003), however this is not surprising given the much lower density of points in each transit signal (15 times fewer). TIC 77951245 is an M4 star of T = 12.375 in the 42 Myr Columba association. Based on the depth of the observed signals and the radius of the star, this signal would correspond to an approximately Jupiter-sized planet if the planetary hypothesis was correct. However, follow-up of this object by the TFOP working group eventually revealed it to be a near-equal mass spectroscopic binary, based on HRS measurements on the 10m SALT telescope on 10th August 2019.\footnote{https://exofop.ipac.caltech.edu/tess/target.php?id=77951245}

These two TOIs along with the new confirmed exoplanet DS Tuc A b discussed in Section \ref{Recovery of DS Tuc A b} represent all three of the TOIs highlighted by the main 2min SPOC pipeline for this selection of young stars. This illustrates the fact that this pipeline is working as effectively as the main SPOC 2min pipeline for these young objects, further emphasising its potential for finding candidate signals around stars present only in the 30min cadence data.

\subsection{Eclipsing Binaries}

\begin{figure*}
	\includegraphics[width =\textwidth]{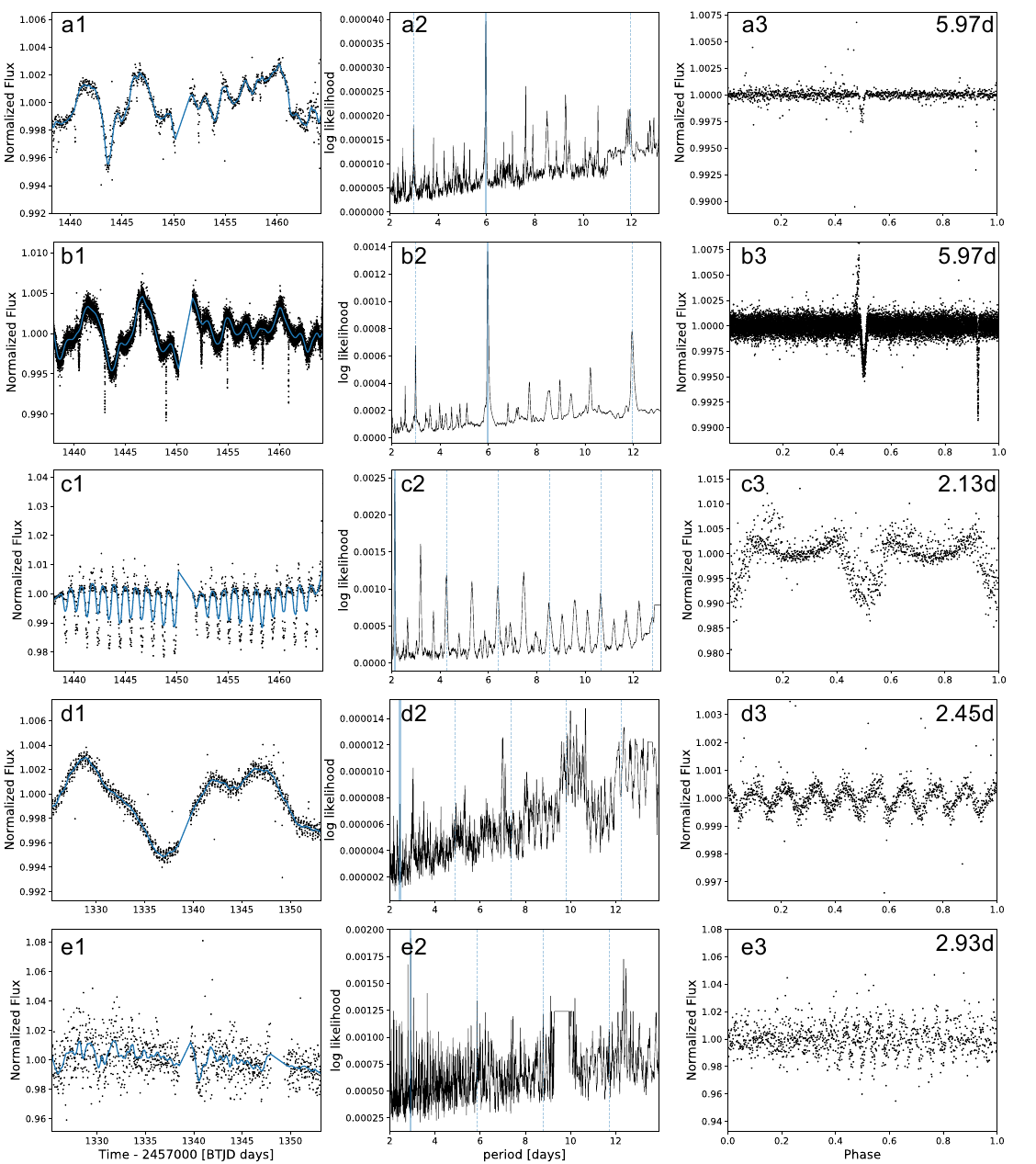}
    \caption{Sector 1-5 light-curve zoo. Column 1 shows the original light-curve for each star with a 20-bin LOWESS-smoothed fit overplotted in blue. Column 2 shows the BLS periodogram for each source after LOWESS-detrending, with the highest power period highlighted in blue (except for J0635-5737 - panel d - where a shorter alias is chosen for clarity), and aliases of this period shown with dotted blue lines. Column 3 shows the flux after 30-bin LOWESS-detrending for each source folded by the highest power period. Individual sources from top to bottom: Panel a: HD 28982, an eccentric EB in ABDMG (approx 150Myr) - DIA 30min data, Sector 5; Panel b: HD 28982 in \textit{TESS} 2min data; Panel c: J0529-2852/TIC 31281820 in COL (approx 42 Myr), demonstrating strongly periodic drops - DIA 30min data, Sector 5; Panel d: J0635-5737/TIC 348839788 in ABDMG (aprox 150 Myr), demonstrating 'double variation' in its rotation - DIA 30min data, Sector 1; Panel e: J0552-5929/TIC 350712873 in THA (approx 45 Myr), demonstrating very fast rotation - DIA 30min data, Sector 1.}
    \label{fig:full_page_lc_table}
\end{figure*}

A number of clear young eclipsing binaries were also revealed by this survey, perhaps the most dramatic of which being HD 28982. This highly eccentric eclisping binary system was revealed in this work by a 30-bin run of the LOWESS-based detrending (without peak cutting) applied to sector 5 of the \textit{TESS} 30min DIA data (see Fig \ref{fig:full_page_lc_table}, panel a). A period of 5.97 days was clearly highlighted as the maximum power period in the BLS periodogram, and two distinctly different duration and depth transit signals were visible. HD 28982 is associated with the AB Doradus Moving Group (ABDMG), which is approximately 150Myr old \citep{Gagne2018BANYAN150pc}. 

Because it has long been known to be a 'bona-fide' member of this moving group it was included in the initial \textit{TESS} CTL \citep{Stassun2019TheList} and received 2min coverage in the main \textit{TESS} survey. This provides an interesting opportunity to compare the 2min and 30min data for this young object. After retrieval from MAST, the 2min data was subjected to the same detrending as the 30min data by expanding the number of bins for each LOWESS-based detrending step to 450, 15 times as many as in the 30min analysis. This revealed the detrended light-curve shown in panel b of Fig \ref{fig:full_page_lc_table}. Aside from the obvious increase in signal to noise and visually more obvious signal provided by the 2min data, the 30min data clearly provides sufficient information to constrain the EB period and durations. However alike to the TOIs, the 30min data suffers in terms of depth accuracy, likely due to exposure smearing from the longer 30min cadence data. In addition, the 30min cadence data from the DIA pipeline is more affected by scatter than the 2min data, likely due to a combination of pointing scatter and improperly corrected scattered light. Care must therefore be taken when evaluating planetary radii for planet candidates based on 30min data alone, and may be better left to higher-cadence photometry in follow-up observations.

\subsection{Rotation and activity} \label{rotation}

Rotation and spurious stellar activity were the cause of the strongest peaks in many of the stars viewed in this sample, which helps to explain why searching for planets around young stars is so much harder than around many older stellar hosts. As a first step in characterisation of this rotation and activity, the period of each star's main stellar variability/activity was also recorded using the generalized Lomb-Scargle method \citep{Lomb1976Least-squaresData} implemented into \texttt{python} as the \texttt{LombScargle} function in the \texttt{python} \texttt{astropy library}\footnote{https://docs.astropy.org/en/stable/api/astropy.timeseries.LombScargle.html} \citep{TheAstropyCollaboration2013Astropy:Astronomy,AstropyCollaboration2018ThePackageb}. Furthermore, the main amplitude of the primary variability was determined in the peak-cutting step discussed in Section \ref{peak_cutting}, using the peaks and troughs identified for removal. A wide variety of rotation and activity curves were observed in this work, broadly separated into four categories: near-uniform periodic, periodic but evolving, aperiodic and fast rotators. Amplitudes of oscillation varied from 0.02\% to 9\%, while primary periods of flux variation varied from 0.122d (e.g. J0552-5929 - panel e of Fig \ref{fig:full_page_lc_table}) to non-variable over the 27-day time-period.

\begin{figure}
	\includegraphics[width=\columnwidth]{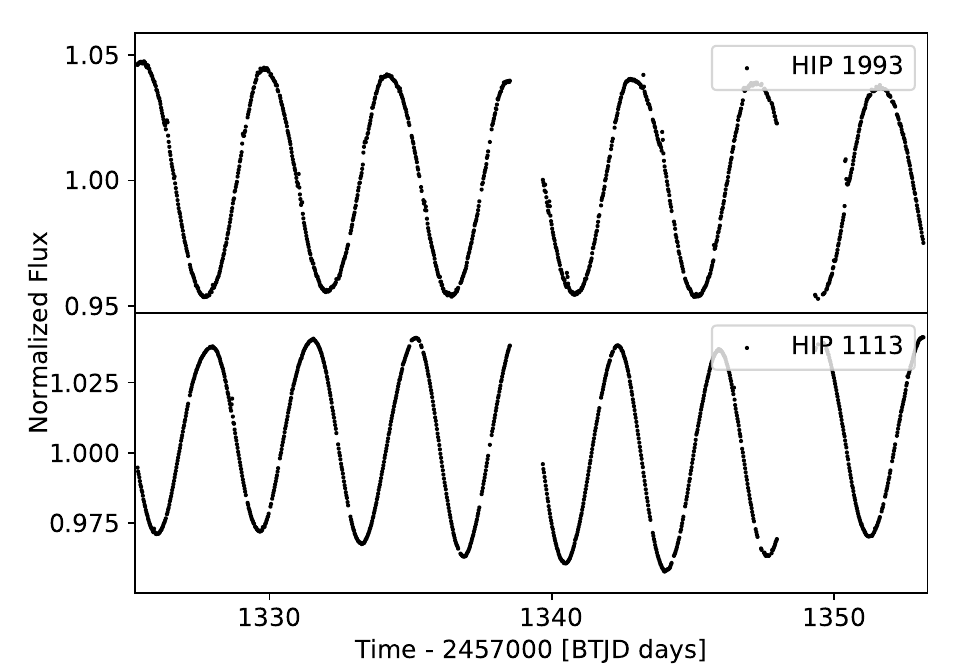}
    \caption{Examples of rotation/activity type 1: Near-uniform sinusoidal variation - HIP 1993 (top) and HIP 1113 (bottom) in Sector 1}
    \label{fig:sinusoidal_var}
\end{figure}

% Type 1: Near-uniform 
The first of these categories, those light-curves which exhibit near-uniform variation in amplitude over a set period, are well illustrated by the sinusoidal curve of sources such as HIP 1993 and HIP 1113 (Fig \ref{fig:sinusoidal_var}), alongside strongly periodic drops in sources like 2MASS 05292529-2852274 (see panel c of Fig \ref{fig:full_page_lc_table}). These types of sources are theoretically ideal for detrending, as they can be easily modelled by quasi-periodic smoothing functions, especially when the period of flux variation is over periods of two or more days. Furthermore, the peak-cutting method discussed in Section \ref{peak_cutting} was often observed to aid the retrieval of injected transits for these periodic sources. However, very fast rotation of this type still provides difficulties, as discussed below.

\begin{figure}
	\includegraphics[width=\columnwidth]{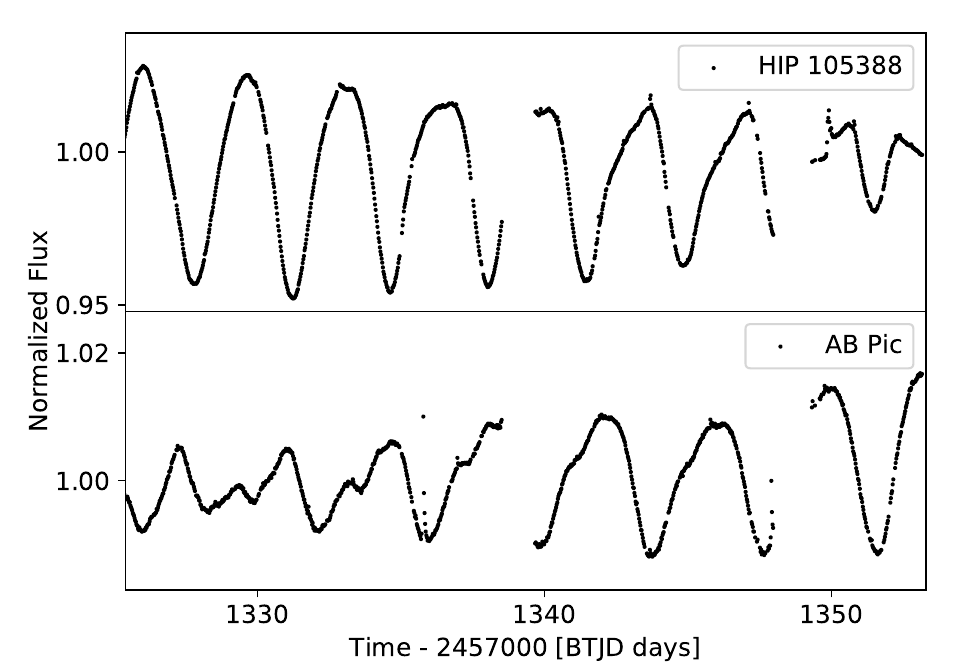}
    \caption{Examples of rotation/activity type 2: Periodic with rapidly evolving amplitudes - HIP 105388 (top) and AB Pictoris (bottom) in Sector 1}
    \label{fig:HIP_105388_AB_Pic}
\end{figure}

%Type 2: Periodic but evolving case - e.g. HIP 105388 and AB Pic:
Another very important category of variation seen are those light-curves with strong periods but obvious variation in flux amplitude. This type of light-curve is well illustrated by sources such as HIP 105388 (observed in \textit{TESS} sector 1) and AB Pic (observed in all southern hemisphere sectors - 1-13) - see Fig \ref{fig:HIP_105388_AB_Pic}. The oscillation amplitudes of these sources are both observed to considerably change over the course of a single sector. Indeed, in the case of AB Pic, the amplitude of the primary oscillations changes from 0.01\% to >0.03\% over the course of the five sectors of data for which the 30min DIA data exists. Other sources such as HIP 1481 appear to exhibit 'beating'-like behaviour in their light-curves, going through periods of more and less intense oscillations yet with similar periods throughout. Thus for these objects, while the strong periodic nature of their oscillations makes the periods easy to identify, blind removal of these varying oscillations is difficult on a wide-scale basis. Nonetheless, the peak-cutting technique discussed in Section \ref{peak_cutting} particularly beneficial for sources of this type, such as HIP 32235 (Fig \ref{fig:Peak_cut_example}) and HIP 105388 (top, Fig \ref{fig:HIP_105388_AB_Pic}).

\begin{figure}
	\includegraphics[width=\columnwidth]{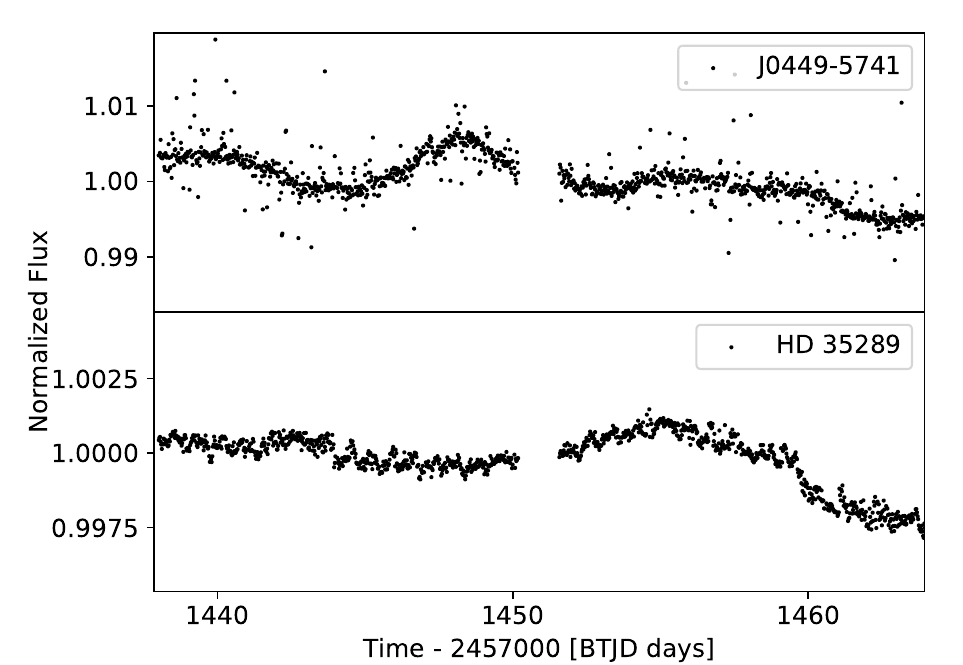}
    \caption{Examples of rotation/activity type 3 - Aperiodic variations. DIA 30min-cadence light-curves for J0449-5741 (TIC 220425740, top) and HD 35289 (bottom) in \textit{TESS} Sector 5}
    \label{fig:Aperiodic}
\end{figure}

%Type 3: Aperiodic:
Alongside these periodic light-curves, a smaller number of aperiodic variations (or perhaps those with rotation or variation periods much longer than the 27 day observation time) were observed. Most of these variations were sufficiently long-period to be easily removed by the base 30-bin LOWESS-smoothing method (as was the case for J0449-5741 and HD 35289 (Fig \ref{fig:Aperiodic})), or are sufficiently aperiodic that any activity would not overcome true periodic transit signals on the BLS periodogram. However, some more complex aperiodic cases exist, such as TIC 348839788 (Panel d, Fig \ref{fig:full_page_lc_table}), where two separate rotation and activity profiles seem to be apparent - one aperiodic larger amplitude evolution clearly evident in the original light-curve, and one much faster rotation-based signal with a period of approximately 0.27 days (see panel d, Fig \ref{fig:full_page_lc_table}. This 'double variation' is present in a small number of other sources too, so should be considered carefully in future detrending efforts. For such sources this will likely require at least two detrending steps (a wider smoothing to remove the large-scale activity, followed by smaller-scale smoothing/modelling of the rapid rotation) in order to yield a flat pre-BLS-search light-curve. 

%Type 4: Fast rotators:
Perhaps the greatest challenge in the search for young exoplanets is the case of rapidly rotating stars such as HIP 22295 (Panel e, Fig \ref{fig:full_page_lc_table}) and CD-46 287 (bottom panel, Fig \ref{fig:Sensitivity_Comparison}. These fast rotators are unfortunately quite common in stars of such young ages (indeed at 52 stars (20.3$\%$) in this sample had rotation periods of less than 1 day), likely due to leftover angular momentum from their formation \citep{Prialnik2009AnEvolution}. For the fastest rotators, 30min data alone struggles to untangle rotation from any potential transit signals, especially when rotation periods of less than 1 day are coupled with significant changes in amplitude. In order to more effectively dissociate such rapid rotation from transit signals, using more sensitive 2min cadence data is preferred, likely coupled with more intelligent modelling than simple smoothing methods. The planned \textit{TESS} 20s cadence data would also aid this effort, as rotation profiles will become more carefully defined and more detail of transit ingresses/egresses may become apparent. However, techniques to overcome this fast rotation challenge still need to be developed. 

\subsection{Sensitivity analysis results}

\subsubsection{Overall results}
\begin{figure}
	\includegraphics[width =\columnwidth]{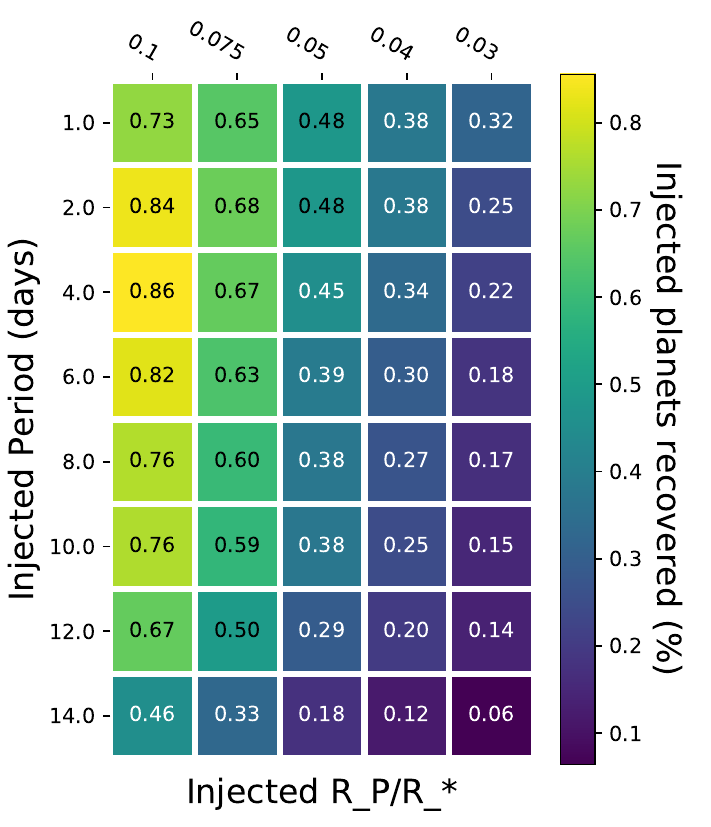}
    \caption{Overall sensitivity analysis results (Sectors 1-5) by injected period and planetary radius. The numerical contents of each box corresponds to the overall percentage recovery of injected planets for each particular combination, with colour-gradient of the same added for clarity.}
    \label{fig:Overall_Heatmap}
\end{figure}

The conducted sensitivity analysis revealed a number of interesting results, for which an overall summary is presented in Table \ref{tab:Sensitivity_overall} and Figure\ref{fig:Overall_Heatmap}. What is immediately apparent is the steep drop-off in recovery as the $R_P/R_*$ radius ratio decreases, falling from 77.6$\%$ at a radius ratio of 0.1 down to 20.4$\%$ at 0.03. This is to be expected as the inherent scatter and leftover variability amplitude of the light-curves steadily overcomes the signal of the smaller injected planets.

\begin{table}
    \addtolength{\tabcolsep}{-2pt}
	\centering
	\caption{Results from complete Sector 1-5 sensitivity analysis for the 256 stars in the BANYAN young star sample with DIA FFI data. Percentage recovery is presented both overall and in each individual sector. Note that results from the 14d injections were excluded from the percentage recoveries due single transits frequently being present at this period. Total number of individual sources for each sector are: Sector 1: 74; Sector 2: 77; Sector 3: 75; Sector 4: 120; Sector 5: 138.}
	\label{tab:Sensitivity_overall}
	\begin{tabular}{ccccccc} % four columns, alignment for each
		\hline
		$R_P/R_*$ & Overall ($\%$) & S1 ($\%$) & S2 ($\%$) & S3 ($\%$)& S4 ($\%$) & S5 ($\%$)\\
		\hline
		0.1   & 77.6 & 79.7 & 80.3 & 73.1 & 69.8 & 84.3\\
		0.075 & 61.7 & 61.6 & 64.0 & 56.8 & 51.4 & 72.2\\
		0.05  & 40.9 & 43.8 & 36.9 & 39.6 & 31.6 & 50.31\\
		0.04  & 30.2 & 31.3 & 30.4 & 31.1 & 20.0 & 38.0\\
		0.03  & 20.4 & 24.3 & 18.4 & 22.1 & 11.1 & 26.6\\
		\hline
	\end{tabular}
\end{table}

Interestingly the recovery of injected signals was not entirely consistent across the different sectors viewed in this analysis. This discrepancy appears to result from two major factors: comparative sector systematics and evolving light-curve amplitudes for a small subset of stars. Regarding the first factor, Sector 4 was the worst affected, with recovery rates for all radius ratios consistently far below the average. This is likely due to the strong reflected light glints seen at the end of each orbit for some sources (particularly those on camera 4), which are harder to systematically remove without also removing useful data from other light-curves. Conversely, recovery was significantly better in Sector 5 compared to the other sectors, with 26.6$\%$ of injected planets still retrieved down to radius ratios of 0.03. This inter-sector variation is well illustrated by the light-curves for J0455-6051/TIC 55651278 (an M5 star in the AB Doradus Moving Group), where injected planetary signals were recovered down to radius ratios of 0.075, 0.05, 0.075, 0.1 and 0.03 in sectors 1, 2, 3, 4 and 5 respectively in the 2-6 day period cases. In this case such variation was caused by a combination of evolving flaring-type stellar variability coupled with extra noise in Sector 4. On the other hand, AB Pictoris (a K1V star in the Carina Association) demonstrates the second source of sector-dependent sensitivity, with injected planetary signals recovered down to steadily larger radius ratios of 0.04, 0.04, 0.075, 0.075 and 0.1 in sectors 1-5 due to the evolving stellar variability increasing in amplitude over time. While differences in the systematics of each sector can reasonably be expected to reduce over time as the \textit{TESS} satellite pointing is refined, the evolution of individually active sources will remain a challenge not only for \textit{TESS} but also for future missions such as \textit{PLATO}.

A slightly more complex trend was observed when varying the injected planetary periods, as is highlighted by the overall heat-map in Figure \ref{fig:Overall_Heatmap}, and the sector-by-sector heat-maps in the Appendix. For all sectors except Sector 2 (where a 2-day period was preferred), the recovery peaked around a 4-day injected period for the deepest (0.1$R_P/R_*$) transits, unlike the shortest 1.0 day period that may be initially expected. However as the injected planet radii decreased a more standard drop in recovery from 1-14d was observed. Such behaviour suggests that this form of analysis is most sensitive to larger planets in the 2-6 day period regime, and most sensitive to closer-orbiting planets as radii decrease further. In case the observed drop for 1-day planets was due to 1-day Earth-related systematics such as Earthshine, the injection/recovery analysis was repeated with a 1.1d period planet, however similar recovery rates were observed to the 1-day case. Instead this discrepancy may be related to a combination of the increased activity of young stars coupled with the small number of data points per transit for such short-period planets.  The primary reason for the much lower recovery of planets with injected periods of 14-days was that data gaps and randomly injected epochs frequently led to single transits appearing in the data-set. This same effect was also observed for some 12-day injections. Some interesting sector-to-sector variations are also clear in the sector-specific heat-maps (see Figures \ref{fig:Sensitivity_S1}-\ref{fig:Sensitivity_S5}), with Sector 4 exhibiting the lowest recovery rates (especially for the smallest radius planets), and Sector 5 the highest.

Another useful product of this sensitivity analysis is the individual sensitivity to planet detection for each star. An overview of this table is presented in Table \ref{tab:Full_Sensitivity_Table}, with a full version available online.

\begin{table*}
    \addtolength{\tabcolsep}{-1pt}
	\centering
	\caption{Full sensitivity analysis table for each of the 256 stars with DIA light-curves. Includes information on the highest likelihood period recovered for each star in every sector it appears, with injected1-14 day period planets and $R_P/R_*$ radius ratios from 0.1 to 0.03. Full table available online.}
	\label{tab:Full_Sensitivity_Table}
	\begin{tabular}{ccccccccccc} 
		\hline
		Target ID & RA (deg) & Dec (deg) & Sector & Injected Period (d) & Radius Ratio & ... & Log likelihood & Max Period (d) & Recovered? & Notes\\
		\hline
		2M0123-6921 & 20.79 & -69.36 & 1 & 1.0 & 0.1 & ... & 0.00121 & 2.65 & TRUE & Alias \\
		2M0123-6921 & 20.79 & -69.36 & 1 & 1.0 & 0.075 & ... & 0.00125 & 3.98 & TRUE & Alias\\
		...  & ... & ... & ... & ... & ... & ... & ... & ... & ... & ...\\
		WX Col B  & 84.30 & -42.72 & 5 & 14.0 & 0.04 & ... & 2.57e-5 & 12.52  & FALSE & - \\
		WX Col B  & 84.30 & -42.72 & 5 & 14.0 & 0.03 & ... & 1.95e-5 & 11.22  & FALSE & -\\
		\hline
	\end{tabular}
\end{table*}

\subsubsection{Rotation period vs recovery}

\begin{figure*}
	\includegraphics[width =\textwidth]{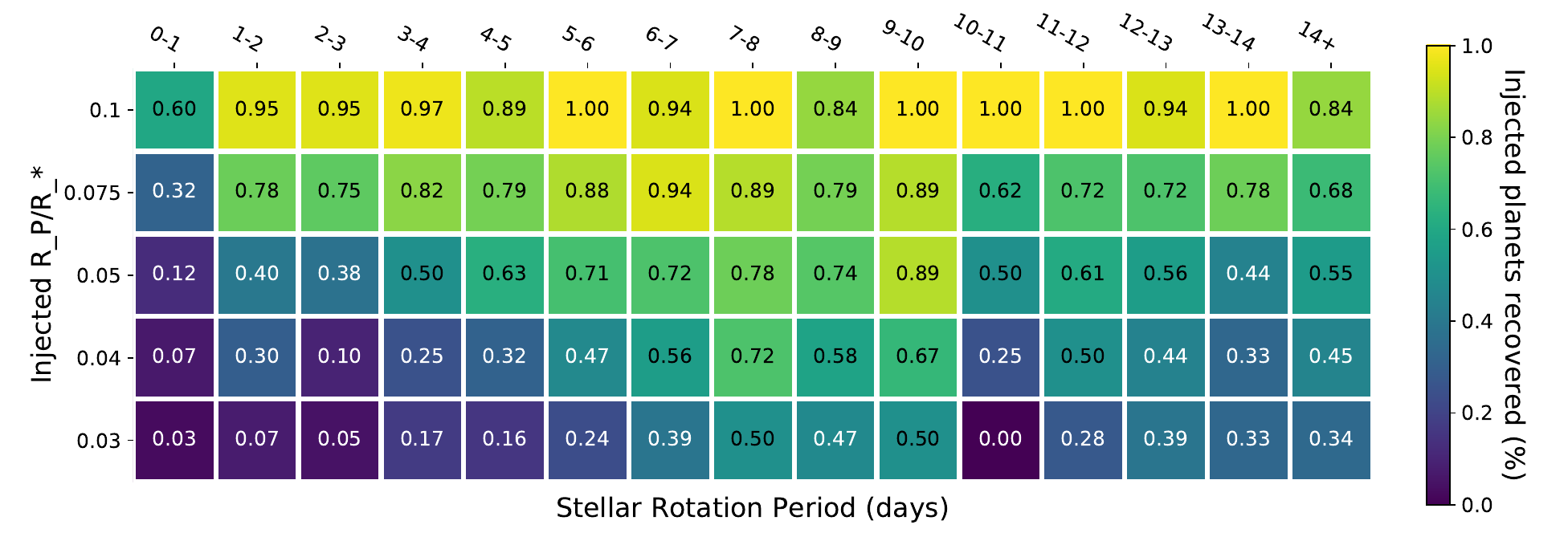}
    \caption{Stellar rotation period vs recovery depth for a 4-day injected period planet across all targets and sectors. Design similar to Figure \ref{fig:Overall_Heatmap}}
    \label{fig:Rot_P_vs_depth}
\end{figure*}

As has been seen in Section \ref{rotation}, a wide range of rotation periods was observed for stars in this sample. This provides an interesting opportunity to test the relationship between rotation rate and recovery, which is especially important given concerns about finding planets around swiftly rotating stars and those with rotation periods near the injected planetary period. In order to investigate this relationship, a injection/recovery analysis of rotation rate vs depth recovered for each target was conducted for each injected period. A typical example of one of these plots is shown in Figure \ref{fig:Rot_P_vs_depth} with an injected period of 4.0 days. Somewhat counter-intuitively, while there is a slight skew towards larger radius ratios below rotation periods of 5 days, there is no significant evidence that the overall recovery depth of injected transits is a function of rotation period, with recoveries down to radius ratios of 0.03 (and complete non-recoveries) across the entire period range. In addition, recovery depth does not seem to be detrimentally affected by being close to the injected period (4 days in this case), as similar recovery was observed for every injected period. Note that many variations between individual period ranges can be explained by the relatively small number of targets in this sample, especially in the 10-11d period case where only eight targets were present with periods in this range. However, it should be noted that 60$\%$ of all unrecovered signals were from targets with rotation periods of less than 1 day. This constituted 40$\%$ of all injected signals into targets with these rotation periods. Furthermore the vast majority (79$\%$) of the recovered light-curves with rotation periods <1 day were only found down to radius ratios of 0.075$R_P/R_*$). This suggests that searching for small planets (<0.075$R_P/R_*$) around stars with rotation periods less than 1 day is likely futile until better techniques are developed to detrend this fast rotation. However, since failures for all other rotation periods were consistent at 0-3 light-curves per 1-day period interval (e.g. 4-5d), searching for planets around stars with rotation periods longer than 1 day shows promise, even down to 0.03$R_P/R_*$ planets.

\subsubsection{Effectiveness of the peak-cutting technique} \label{Peak_cutting_comparison}

Since the peak-cutting technique exhibited variable effectiveness according to the shape of the light-curves in building the pipeline, much of the prior analysis was undertaken without the peak-cutting option applied. However, to test the wider effectiveness of this option and evaluate where its application was most useful, a comparison study was undertaken for all targets in Sector 1, both with and without peak-cutting. For this comparison test planets with a set period of 8.0d and radius ratios of $R_P/R_* = 0.1-0.03$ were injected into each of the light-curves in Sector 1. Light-curves were then detrended using the standard 30-bin LOWESS-detrending method described above and searched through using the standard BLS method.

Of the 74 Sector 1 targets in the sample with light-curves available from \citet{Oelkers2018PrecisionApproach}'s DIA FFI pipeline, the basic peak-cutting technique failed for 13-15 of the objects (depending on the injected radius ratio). In all cases this was due to peaks and/or troughs not being located by the automatic \texttt{find\_peaks} function, typically because the light-curves were simply too flat to exhibit any significant peaks or troughs. Of the remaining 59-61 light-curves, recovery with peak-cutting was in general comparable, or slightly better than, the recovery of planets when peak-cutting wasn't applied (as is summarised in Table \ref{tab:Peak_cutting_comparion}). The one exception to this was the 0.1$R_P/R_*$ case, which was caused by the peak-cut analysis failing for three of the original 0.1 radius ratio light-curves.

\begin{table}
	\centering
	\caption{Comparison between percentage recovery of injected planets in Sector 1 for both in the original and peak-cut light-curves.}
	\label{tab:Peak_cutting_comparion}
	\begin{tabular}{cccc} % four columns, alignment for each
		\hline
		Radius Ratio & Number of lcs & Original ($\%$) & Peak-Cut ($\%$) \\
		\hline
		0.1   & 61 & 90.2 & 82.0\\
		0.075 & 59 & 62.7 & 64.4\\
		0.05  & 59 & 39.0 & 40.7\\
		0.04  & 60 & 31.7 & 35.0\\
		0.03  & 60 & 25.0 & 25.0\\
		\hline
	\end{tabular}
\end{table}

However, it is when looking at individual targets that the power of the peak-cutting technique is most evident. For seven of the targets in Sector 1, the use of the peak-cutting technique yielded significant improvements in the recovery of smaller injected radius ratios. The most significant improvements were seen for the object J0247-6808, where injected planets were recovered down to a radius ratio of 0.03$R_P/R_*$, despite not being recovered for any radius ratio in the non-peak-cutting case. Similarly in the case of HIP 32235 the depth recovered dropped to 0.03 from 0.075 in the non-peak-cut analysis. Similar improvements (though with less significant drops in recovered depth) were observed for RBS 38, TYC 8895-112-1, J0346-6246, J0414-7025 and J2231-5709.

There were two different reasons why recovery was improved for these objects - one showing the technique working as designed and the other a fortuitous side-effect. The former can be seen in HIP 32235, RBS 38, TYC 8895-112-1 and J2231-5709 where each light-curve exhibits sharp variability peaks with periods of order 3-8 days and amplitudes greater than 2.5$\%$. In this case the technique aids recovery of the planets by successfully cutting the sharp turning points of these light-curves which were previously leftover as false-transits after the LOWESS-smoothing step. It is these types of light-curve variablity (approximately $3 < P_{rot} < 8$ days; Amplitude $> 2.5\%$) which are best-handled by using the peak-cutting technique. All light-curves with rotation periods of 3-8 days which weren't improved by peak-cutting were later found to have had at least one of their peaks cut, however their depths reached were not affected.

The other three light-curves improved by the use of this technique (J0247-6808, J0346-6246 and J0414-7025) were aided accidentally, having sections of increased scatter masked as a result of the applied cuts. This is a convenient side-effect of searching for planets among light-curves with intrinsically higher scatter, but less scientifically interesting.

The limitations of this technique were identified by investigating the small selection of stars detrimentally affected by the peak-cutting. The three main failure modes were:

\begin{enumerate}
    \item Effectively flat light-curves dominated by scatter
    \item Light-curves with activity/variablity periods of $\sim$ 2 days
    \item HIP 33737 - a star with a flat-bottomed rotation activity
\end{enumerate}

In the first case, all peak-cut does is remove useful data, since no significant activity-based peaks and troughs were present. Meanwhile in the 2-day activity/variability period case, a significant portion of the data is cut, with the remaining intervals between each cut too short to be effectively flattened by the 20-30bin (10-15hr) LOWESS-smoothing. These two failure modes effectively represent the two limits of usefulness for this technique, of order 15 days and 2 days respectively. The final case (HIP 33737) was a unique one, where the light-curve exhibited unusually flat-bottomed troughs well-handled by the LOWESS-smoothing technique, and thus the peak-cutting only removed useful data, alike to the first failure mode.

Crucially however, for all light-curves which did not fall into one of the three failure modes identified above, the use of the peak-cutting technique was not found to affect the shallowest depth of transit recovered. This alleviates the chief concern about the use of the peak-cutting technique: that the recovery depth may be reduced if transits near peaks and troughs are inadvertently cut. Overall then, this peak-cutting technique shows the greatest effectiveness for stars exhibiting activity/variability of periods 3-8 days, but may be applied to all stars with periods of approximately 3-15 days without significant detriment, unless those light-curves are particularly flat.

%Based on the sensitivity analysis performed for this sample and the fact that only two exoplanets were found (AU Mic b and DS Tuc A b) the planet occurrence rate for exoplanets with radius ratios down to 0.03 $R_P/R_*$ around young stars is $<$... with 95$\%$ confidence. However, this value should be treated with some caution due to the detrending pipeline struggling with sharp flux variations for many of the sources with activity periods of less than 1 day. A wider population study of exoplanets around young stars will be the focus of a future paper. 
% This planetary occurrence sentence needs a bit of work... Maybe leave for a future paper.
% 2 planets in 256 sources, with 59.4\% sensitivity/recovery on average.
% Also consider geometric likelihood of detection?

%%%%%%%%%%%%%%%%%%%%%%%%%%%%%%%%%%%%%%%%%%%%%%%%%%%%%%%%%%%%%%%%%%%%%%%%%%%%%%%%%%%%%%%%%%%%%%%%%%%%%%

\section{Discussion} \label{Discussion}

As this work has shown, young host stars present many extra challenges in comparison to the generally older, less active stars previous exoplanetary searches have been biased towards. Quicker rotation, increased amplitude activity and other strong stellar-based periodicities wash out candidate planetary signals in BLS searches and make finding transiting signals harder using the traditional automated exoplanet-searching tools. Furthermore, by eye the large amplitude variation of many sources such as HIP 105388 effectively hide real transit signals unless careful detrending is applied first. 

Nonetheless, the methods presented in this work have shown promise at pushing down to lower-radius planets around young stars. The base LOWESS-detrending method provides a useful combination of smoothing and polynomial fitting and, as demonstrated by \citep{Hippke2019WotanPython}, generally outperforms more traditional exoplanet smoothing methods such as Savitzky-Golay filters for younger stars due to its weighted approach to smoothing. Furthermore, for sharp but evolving oscillations such as those seen for HIP 33235, cutting the peaks and troughs of these oscillations yields a significant improvement in the recovery of injected planets, especially for periods of 3-8 days. After initial recovery, the shape and depth of the retrieved transit is then greatly improved by incorporating the developed activity interpolation over transit gaps. 

\begin{figure}
	\includegraphics[width = \columnwidth]{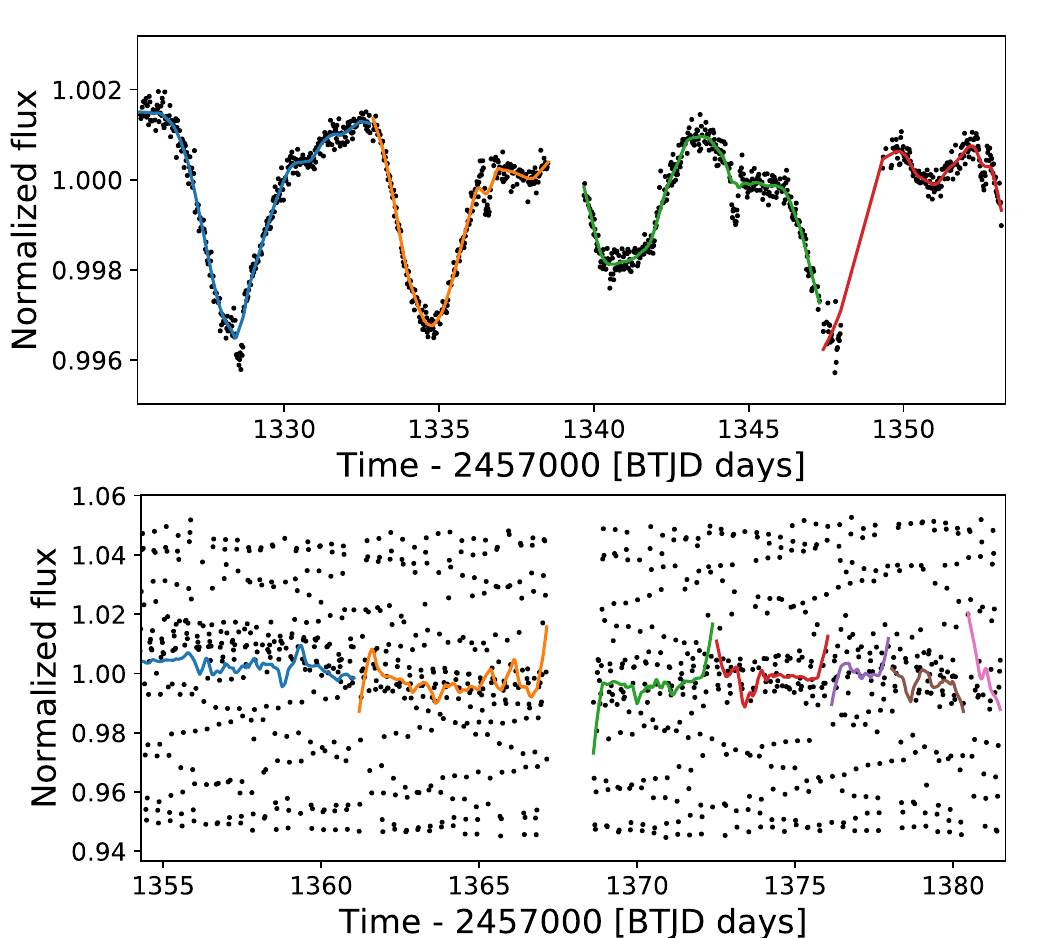}
    \caption{Light-curve comparison for HD 202969 (above) and CD-46 287 (below) with an 8.0d, 0.03 $R_P/R_*$ radius ratio injected planet. While the LOWESS-detrending pipeline has no trouble modelling HD 202969 due to its relatively long duration variation, it struggles to detrend the sharp 0.37d period variability of CD-46 287.}
    \label{fig:Sensitivity_Comparison}
\end{figure}

However, as demonstrated by the sensitivity analysis, the large range in activity type, period and amplitude coupled with the variation in scatter for each source results in significantly different recovery rates for injected planets overall, ranging from sources like CD-46 287 (a K6Ve star in the 35 Myr Octans association, viewed in sector 2) where no planet was recovered even for the largest radius ratio, to HD 202969 (an F8/G0V star also from Octans, viewed in sector 1), where the injected planet was easily recovered across the whole radius ratio range. Comparing these two light-curves (shown in Fig \ref{fig:Sensitivity_Comparison}), the reason for this discrepancy is immediately clear; while the period of the rotational variation in the flux from HD 202969 is much longer than that of the transit duration (and easy to smooth with the 30-bin LOWESS-detrending technique), the rotation period of CD-46 287 is 0.37d, well within the realm of the duration of a planetary transit. Since this pipeline is designed to only remove variation with periods longer than a transit it struggles to remove such short period variations, and as such cannot recover injected planets from such light-curves. 

Observing other sources for which no injected planet sizes were ever recovered, the major challenges appear to be rotation with periods < 1 day, excess scatter, large amplitude flares or other spurious outlying points in the light-curve. These latter two problems present a problem to automatic detrending due to their unique light-curve by light-curve behaviour. Nonetheless these could conceivably be flagged by shape or sigma-clipping and handled on a case-by-case basis. On the other hand, scatter falls into two categories: time-localised scatter such as the scattered light glints seen at the end of some sector 4 light-curves, and non-localised scatter which affects the entirety of the light-curve. The first of these problems is soluble, and is likely best handled in a similar way to flares and instrumental scatter, by flagging and removing specific time periods which are affected. However, excess scatter which extends to the full extent of the light-curve (which can be in excess of 2\%, as in the case of J0122-2566) likely precludes the use of these light-curves for planet searches. This latter type of scatter was most common for very dim stars in the sample, largely with \textit{TESS} T magnitudes in excess of 13. 

The problem of fast rotation is a more systematic issue and is arguably the most important detrending-related challenge presented by young stars. In this work the detrending pipeline struggled to recover injected planetary signals in light-curves with large amplitude flux changes coupled with rotation periods of less than 1 day, largely due to the width of the LOWESS-smoothing being at least 20-bins, or 10hrs wide. Since any variability with a period of 1 day or less will involve at the very least one turning point, such widely-spaced windows can not reasonably be expected to accurately trace and smooth such activity/rotation. Indeed, as discussed in section \ref{rotation}, some of the fastest rotation observed had periods of less than 0.4 days, or less than 10hrs. However, it is not unheard of for verified planets such as Kepler-1283 b \citep{Morton2016FALSEPOSITIVES} to have transit durations as long as 0.4 days, so dropping the LOWESS-smoothing window to any shorter than this could result in significant distortion of potential transit shapes along with removal of stellar activity signals. It is thus clear that a more targeted method is required for such fast rotators, likely with more intelligent modelling-based methods such as Gaussian Processes \citep[e.g.][]{Gillen2020NGTS1}. Unfortunately however GP methods are much more intensive than a simple smoothing or polynomial based methods, and require a more informed knowledge of a star's characteristics. This makes GP-based techniques very well suited to in-depth followup of candidates, but perhaps applying such methods to the entire stellar sample would be less efficient than simply flagging them in an initial BLS periodogram and analysing them separately. At these early stages it is still worth considering other modelling methods as well (such as fitting a sum of sinusuoids as in \citet{Gillen2020NGTS1}), especially since \citet{Hippke2019WotanPython} found such a range in effectiveness in young star detrending methods. If a simpler algorithmic method can be found to model or detrend these very fast rotators then it could help to speed up wide-field transit searches around young, rapidly rotating stars.

Another interesting challenge highlighted by the sensitivity analysis is the rapid evolution of some young star light-curves. This evolution makes modelling the light-curves more difficult and also means that detrending is more effective in some regions of the light-curve than others. As demonstrated for AB Pictoris, this changes how easy it is to recover injected (or no doubt real) exoplanet signals depending on when in the activity cycle one views the star. If however one gains a better understanding of a star's activity cycle through long-term monitoring and asteroseismology, this evolving activity can become an opportunity for increasing the effectiveness of planet searches by targeting quiet sections of the stellar activity cycle. This knowledge would be crucial for radial velocity follow-up of young host stars, since as \citet{Oshagh2017UnderstandingMeasurements} have shown, for very active stars radial-velocity jitter is  correlated with photometric variation in stellar light-curves. Hence being able to predict epochs of low stellar activity based on knowledge gained from photometric monitoring of these young stars will prove crucial for characterising any discovered planets through radial-velocity followup. 

One thing that was made quite clear from this initial survey of young star light-curves is that the 'one size fits most' approach of large-scale photometric surveys such as \textit{Kepler}, \textit{K2} and the main \textit{TESS} SPOC pipeline is often not appropriate for young stars given the large range in shape, amplitudes and periods of light-curve variability observed. Indeed, as discussed in section \ref{rotation}, even in this relatively small number of young stars surveyed periods were observed to vary between 0.27d to non-variable over the 27d observation window. Meanwhile activity and rotation behaviour varied from near-uniform to constantly evolving, and variability shape changed from beating sinusoids to almost flat aside from significant flaring activity. In order to more comprehensively search for planets around such active stars, it is recommended that future detrending pipelines focus on more effectively matching detrending techniques to the primary type of light-curve variability observed. This could be achieved through an initial automated variability-type assignment, similar to - but more advanced than - the current 'variable' vs 'non-variable' assignment implemented into the \textit{Kepler/K2/TESS} pipelines. By creating defined groups of similarly shaped light-curves and types of variability/activity, machine-learning techniques could then be used to assign light-curves the most effective detrending technique based on their  perceived 'group' of variability. Particularly important will be separating known types of intrinsic variability, quick rotators and rapidly evolving light-curves, however many other important groups may become obvious with time. A more in-depth look into different types of variability in young stars over all sectors would thus be very beneficial, and may inform future detrending methods in missions like \textit{PLATO} \citep{Rauer2014TheMission}.

Attempting to understand these stars in more detail raises the important question of whether 30min data is enough to characterise any discovered variability or exoplanet candidate signals, or whether 2min cadence data is required. In this work 30min cadence light-curves were shown to effectively find the period of primary variability (and often second and third variability periods), and to recover all of the currently known TOIs identified through the 2min data, so it is undeniably a very powerful data source for the general stellar sample. However, for those TOIs recovered, exposure smearing resulted in smaller transit depths and thus less accurate transit parameters compared to the 2min data. This highlights the importance of the 2min data (or alternative follow-up photometry) for accurate characterisation of any discovered signals. Furthermore, while the 30min data was sufficient for characterisation of longer-period variability and activity in the light-curves, as the community pushes towards shorter period rotation (especially that with periods of less than 1 day), dissociating transit signals from stellar activity and rotation signals becomes increasingly difficult. This is the realm where 2min data may be crucial in the search for young exoplanets. Furthermore, as the community attempts to understand the causes and evolution of activity in young stars, 2min cadence data would significantly aid asteroseismological efforts for these young stars. However, a full comparison between 2min and 30min data-sets still needs to be undertaken before significant conclusions can be made. 

One further major stumbling-lock remains in the way of finding significant numbers of exoplanets around young stars: so far, a large enough sample of 'bona-fide' young stars does not exist. Even considering basic transit statistics, only approximately 0.47\% of sun-like stars are expected to host Hot Jupiters \citep{Haswell2010TransitingExoplanets}, so considering that only 3076 stars exist in the extended BANYAN sample (some of which will not even be viewed by \textit{TESS}'s primary mission), at most 15 such planets could be expected to be found around these young stars. \citet{Bouma2019Cluster7} attempted to remedy this situation somewhat by concatenating 13 different catalogs of young stars and cluster members from literature, yielding 1,061,447 individual target stars. However, because of the mix of catalogs used and the non-homogenous membership criterion applied, this may include some stars that are not truly young. Furthermore, because it contains many members from within clusters, \textit{TESS} will struggle to create accurate light-curves for many of these stars due to blending on its relatively large 21" pixels. Another promising method that has come into its own with the release of a significant amount of data from the \textit{Gaia} satellite \citep{GaiaCollaboration2016TheMission,GaiaCollaboration2018GaiaProperties} has been expanding cluster and association membership through proper- and galactic-motion relationships. Groups such as \citet{Damiani2019StellarData} and \citet{Lodieu2019APopulation} have used these effectively already to expand the Sco-OB2 and Hyades associations, however each of these stars still needs an independent sign of youth before it can be accepted as a 'bona-fide' member of a stellar association. This independent sign of youth is currently the biggest bottle-neck in expanding the known sample of young stars (especially since many confirmation methods require individual spectroscopic follow-up), so searching for new signs of youth in photomtery or astrometry has the potential to have a significant impact.

In summary, although searching for planets around young, active stars undeniably presents several extra challenges compared to searching around older host stars, the techniques developed in this and other recent work on young stars are beginning to delve more effectively into this age range. However, two major stumbling-blocks remain before the search for young exoplanets can come into fruition: the relatively small number of bona-fide young stars and the very fast rotation of many young stars. Very important groundwork is thus still required to increase this sample of known young stars, and to develop techniques which can more effectively probe young potential host stars with shorter period stellar variability.

\section{Summary} \label{Conclusions}

In this work techniques have been developed to aid the search for transiting exoplanets around young, active stars in the 30min cadence \textit{TESS} FFI data. Young exoplanets (<1Gyr in age) inhabit a very important part of the exoplanet evolutionary timescale, where formation mechanisms, accretion, migration and dynamical interactions can significantly change the shape of observed planetary systems. However, they are also typically situated around young, active and often rapidly rotating host stars, severely hampering the discovery of new planets using the transit method. The developed method attempts to detrend these spurious stellar activity signatures using a 20-30 bin LOWESS method of \citet{Cleveland1979RobustScatterplots} at its base, combined with automated peak-cutting and activity interpolation over transit gaps in order to more effectively differentiate activity from transit signals and preserve the transit shape. A basic version of this pipeline is made available online.\footnote{https://github.com/mbattley/YSD} It is hoped that using this method in tandem with other detrending/transit-search pipelines such as the main \textit{TESS} SPOC pipeline \citep{Jenkins2016TheCenter} and Gaussian-Process based methods \citep[e.g.][]{Gillen2020NGTS1} will expand the number of young planets that can be found.
%combining weighted smoothing with polynomial fitting over a window size of 20-30 bins (10-15hrs). For the sharpest variability an automatic 'peak-cutting' method is then applied, in order to more effectively differentiate between transit signals and sharp periodic activity signals. Transit recovery is then found to be improved further via light-curve interpolation over masked transit signals.

These techniques were applied to young stars in stellar associations from the extended BANYAN sample \citep{Gagne2018BANYAN150pc,Gagne2018BANYAN.Data,Gagne2018BANYAN.2}, using the \textit{TESS} sector 1-5 light-curves derived from the Difference Imaging Approach (DIA) pipeline of \citet{Oelkers2018PrecisionApproach}. Lacking the data quality-flags of the \textit{TESS} 2min cadence data, periods of excess pointing scatter were instead removed by considering the \textit{TESS} data release notes and the engineering quaternion data.  

While no new exoplanet candidates were revealed in this work, results from this initial survey revealed a variety of interesting objects, including the retrieval of the new young exoplanet DS Tuc Ab, TOI 447.01, TOI 450.01, a number of eclipsing binaries and a large array of interesting rotation and activity. In order to test the sensitivity of the developed detrending techniques to different planetary sizes, model \texttt{batman} transits \citep{Kreidberg2015Batman:Python} were injected into each of the young star light-curves at a range of $R_P/R_*$ radius ratios from 0.1 to 0.03 and periods from 1-14 days. The percentage of recovered transit signals from the injected planets dropped from 77.6\% at a radius ratio of 0.1$R_P/R_*$ to 20.4\% at a ratio of 0.03, however was seen to vary considerably between different targets and sectors. Meanwhile while increasing the injected planet period was seen to result in a decreasing recovery rate for smaller planets, the recovery rate was actually observed to peak around periods of 2-6 days for larger planets. An investigation into the relationship between rotation period and recovery depth did not suggest that the two were significantly correlated, aside from a slight skew towards larger planets at short rotation periods and the known difficulty of very short-period (<1day) rotation.

These results alongside deeper examination of light-curves in this sample lead to a number of interesting conclusions. The sensitivity of specific young star light-curves to transit searches appears to be most limited by fast rotation (<1 day rotation periods), excess scatter, scattered light glints and significant flaring activity. Meanwhile the rapid evolution of many young star light-curves offers both a detrending challenge and a potential opportunity to search more efficiently for exoplanets at less active times. Given the vast array of different types of young star light-curves seen in this initial survey, in the future a multi-faceted detrending approach which first classifies light-curves according to their broad activity/variability is perceived as beneficial. It is clear that the 30min cadence data shows particular promise for additional detections to the main SPOC 2min pipeline, as it is capable of retrieving all of the TOIs highlighted in the sectors analysed. However, planet parameters derived from the 30min light-curves can be less reliable due to phenomena such as exposure smearing. Thus the acquisition of 2min light-curves for young stars remains desirable, especially since this will also aid in-depth astroseismic characterisation of these sources.

Overall, the two largest challenges remaining before significant progress can be made in the field of young exoplanet science are those of very rapidly rotating stars and the relatively small numbers of confirmed young stars. Solving these two problems has the potential to gift the exoplanetary community with a significant advancement in understanding of how exoplanets form and develop into stable exoplanetary systems.

\section*{Acknowledgements}

%The Acknowledgements section is not numbered. Here you can thank helpful colleagues, acknowledge funding agencies, telescopes and facilities used etc.

The authors would like to thank the anonymous referee for their comments which improved the quality and robustness of this paper.

This paper includes data collected by the \textit{TESS} mission. We acknowledge the use of public TOI Release data from pipelines at the \textit{TESS} Science Office and at the \textit{TESS} Science Processing Operations Center. Funding for the \textit{TESS} mission is provided by NASA’s Science Mission directorate. This research has made use of the Exoplanet Follow-up Observation Program website, which is operated by the California Institute of Technology, under contract with the National Aeronautics and Space Administration under the Exoplanet Exploration Program. \textit{TESS} data were obtained from the Mikulski Archive for Space Telescopes (MAST). STScI is operated by the Association of Universities for Research in Astronomy, Inc., under NASA contract NAS5-26555. Support for MAST is provided by the NASA Office of Space Science via grant NNX13AC07G and by other grants and contracts. This research has made use of the NASA Exoplanet Archive, which is operated by the California Institute of Technology, under contract with the National Aeronautics and Space Administration under the Exoplanet Exploration Program. MPB acknowledges support from the University of Warwick via the Chancellor's International Scholarship. DLP acknowledges support from STFC and also the Royal Society. DJA acknowledges support from the STFC via an Ernest Rutherford Fellowship (ST/R00384X/1).

%%%%%%%%%%%%%%%%%%%%%%%%%%%%%%%%%%%%%%%%%%%%%%%%%%



\bibliographystyle{mnras}
\bibliography{references} % if your bibtex file is called example.bib

%%%%%%%%%%%%%%%%%%%%%%%%%%%%%%%%%%%%%%%%%%%%%%%%%%

%%%%%%%%%%%%%%%%% APPENDICES %%%%%%%%%%%%%%%%%%%%%

\appendix

\section{Sensitivity Analysis for each Sector}

A sector by sector breakdown for the conducted sensitivity analysis is given below in Figures \ref{fig:Sensitivity_S1}-\ref{fig:Sensitivity_S5}. Figure format after that explained in Figure \ref{fig:Overall_Heatmap}.

\begin{figure}
	\includegraphics[width =\columnwidth]{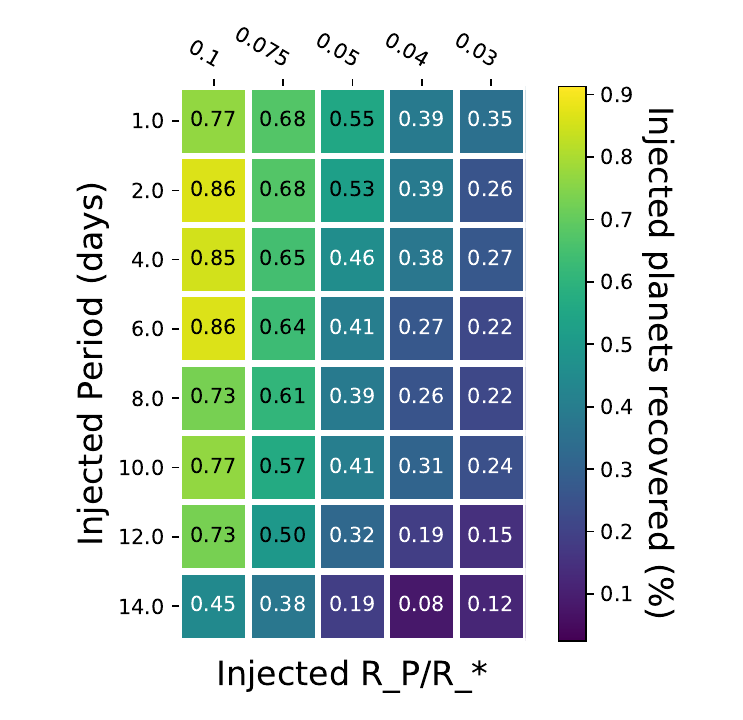}
    \caption{Sensitivity Analysis for Sector 1. Total number of sources = 74}
    \label{fig:Sensitivity_S1}
\end{figure}

\begin{figure}
	\includegraphics[width =\columnwidth]{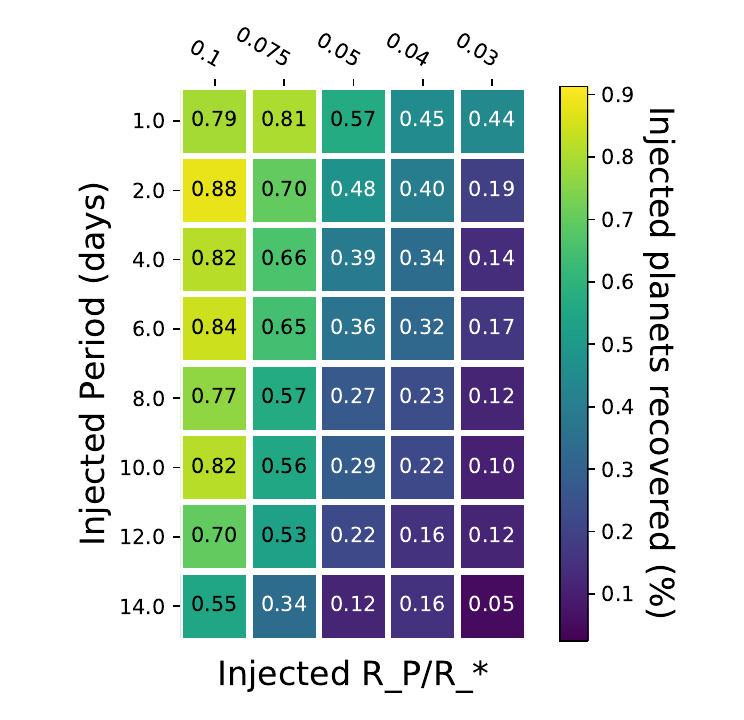}
    \caption{Sensitivity Analysis for Sector 2. Total number of sources = 77}
    \label{fig:Sensitivity_S2}
\end{figure}

\begin{figure}
	\includegraphics[width =\columnwidth]{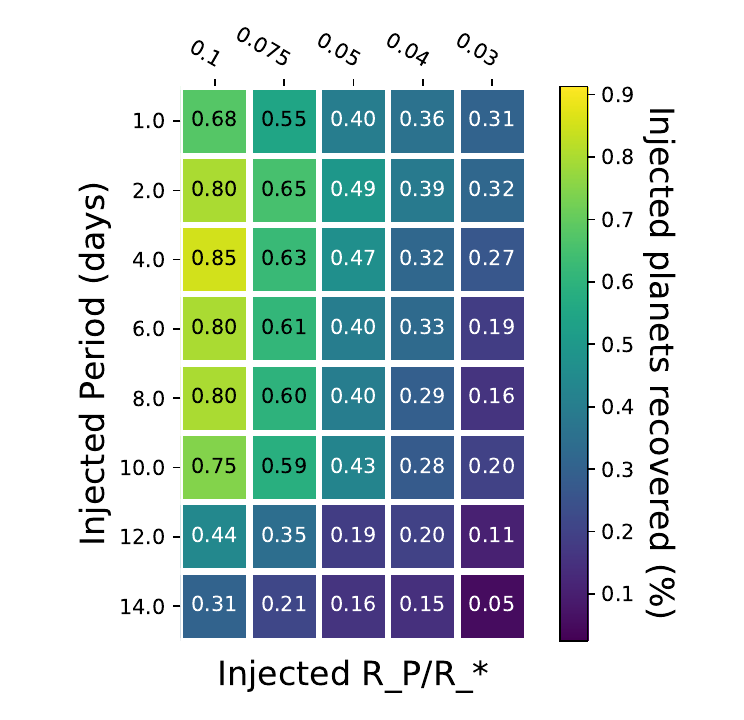}
    \caption{Sensitivity Analysis for Sector 3. Total number of sources = 75}
    \label{fig:Sensitivity_S3}
\end{figure}

\begin{figure}
	\includegraphics[width =\columnwidth]{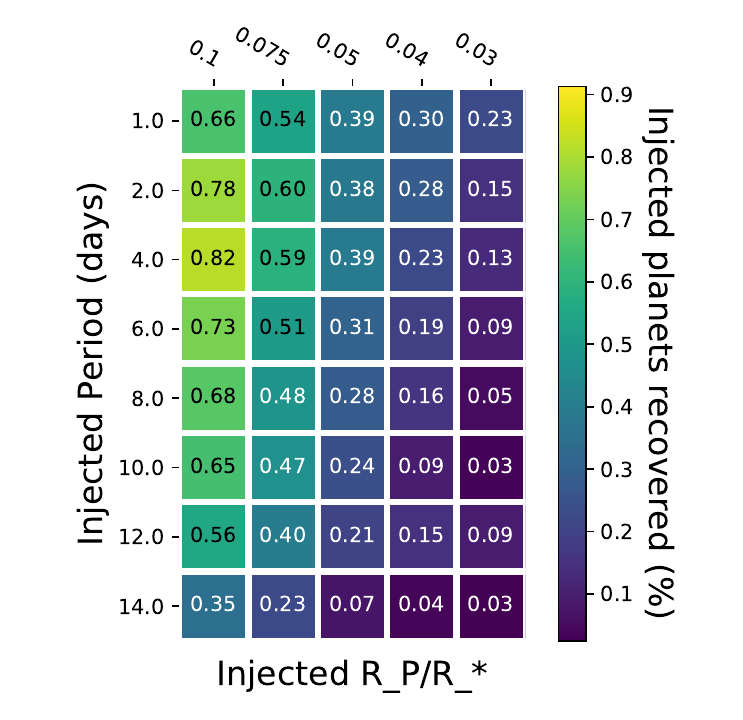}
    \caption{Sensitivity Analysis for Sector 4. Total number of sources = 120}
    \label{fig:Sensitivity_S4}
\end{figure}

\begin{figure}
	\includegraphics[width =\columnwidth]{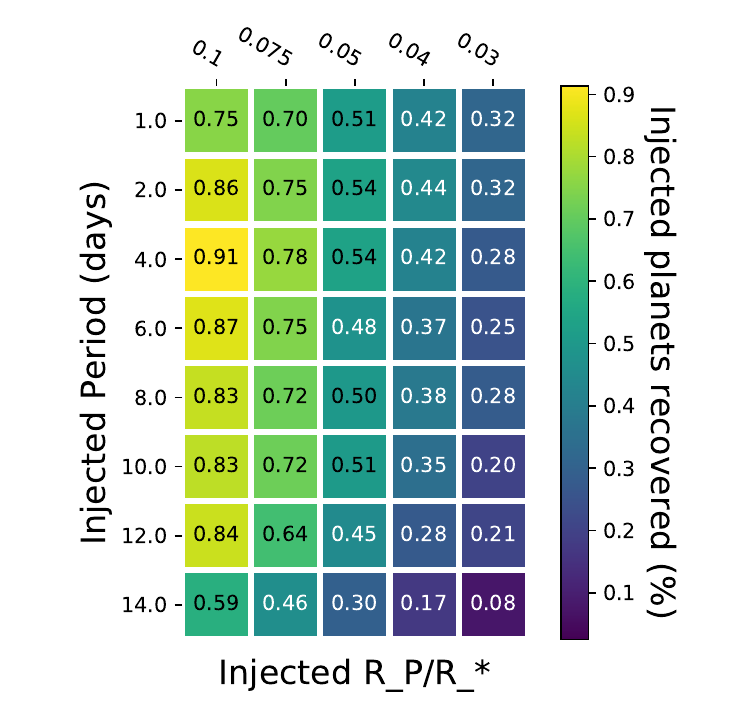}
    \caption{Sensitivity Analysis for Sector 5. Total number of sources = 138}
    \label{fig:Sensitivity_S5}
\end{figure}

%%%%%%%%%%%%%%%%%%%%%%%%%%%%%%%%%%%%%%%%%%%%%%%%%%

% Don't change these lines
\bsp	% typesetting comment
\label{lastpage}
\end{document}